\documentclass[twocolumn]{aastex631}

\submitjournal{AJ}

\shorttitle{High-resolution spectroscopy of the terrestrial exoplanet GJ 486b}
\shortauthors{A. Ridden-Harper et al.}
\graphicspath{{./}}

\begin{document}

\title{High-Resolution Transmission Spectroscopy of the Terrestrial Exoplanet GJ 486b}

\correspondingauthor{Andrew Ridden-Harper}
\email{ariddenharper@lco.global}

\author[0000-0002-5425-2655]{Andrew Ridden-Harper}
\affiliation{Department of Astronomy and Carl Sagan Institute, Cornell University, Ithaca, New York 14853, USA}
\affiliation{Las Cumbres Observatory, 6740 Cortona Drive, Suite 102, Goleta, CA 93117, USA}

\author[0000-0003-4698-6285]{Stevanus K. Nugroho}
\affiliation{Astrobiology Center, 2-21-1 Osawa, Mitaka, Tokyo 181-8588, Japan}
\affiliation{National Astronomical Observatory of Japan, 2-21-1 Osawa, Mitaka, Tokyo 181-8588, Japan}

\author[0000-0001-6362-0571]{Laura Flagg}
\affiliation{Department of Astronomy and Carl Sagan Institute, Cornell University, Ithaca, New York 14853, USA}

\author[0000-0001-5349-6853]{Ray Jayawardhana}
\affiliation{Department of Astronomy, Cornell University, Ithaca, New York 14853, USA}

\author[0000-0001-7836-1787]{Jake D. Turner}
\affiliation{Department of Astronomy and Carl Sagan Institute, Cornell University, Ithaca, New York 14853, USA}
\affiliation{NHFP Sagan Fellow}

\author[0000-0001-6391-9266]{Ernst de Mooij}
\affiliation{Astrophysics Research Centre, School of Mathematics and Physics, Queen’s University Belfast, University Road,
Belfast BT7 1NN, United Kingdom}

\author[0000-0003-4816-3469]{Ryan MacDonald}
\affiliation{Department of Astronomy and Carl Sagan Institute, Cornell University, Ithaca, New York 14853, USA}

\author[0000-0001-9796-2158]{Emily Deibert}
\affiliation{David A. Dunlap Department of Astronomy \& Astrophysics, University of Toronto, Toronto, ON M5S 3H4, Canada}
\affiliation{Dunlap Institute for Astronomy \& Astrophysics, University of Toronto, Toronto, ON M5S 3H4, Canada}
\affiliation{Gemini Observatory, NSF’s NOIRLab, Casilla 603, La Serena, Chile}

\author[0000-0002-6510-0681]{Motohide Tamura}
\affiliation{Department of Astronomy, Graduate School of Science, The University of Tokyo, 7-3-1 Hongo, Bunkyo-ku, Tokyo 113-0033, Japan}
\affiliation{Astrobiology Center, 2-21-1 Osawa, Mitaka, Tokyo 181-8588, Japan}
\affiliation{National Astronomical Observatory of Japan, 2-21-1 Osawa, Mitaka, Tokyo 181-8588, Japan}

\author[0000-0001-6181-3142]{Takayuki Kotani}
\affiliation{Astrobiology Center, 2-21-1 Osawa, Mitaka, Tokyo 181-8588, Japan}
\affiliation{National Astronomical Observatory of Japan, 2-21-1 Osawa, Mitaka, Tokyo 181-8588, Japan}
\affiliation{Department of Astronomical Science, The Graduate University for Advanced Studies, SOKENDAI, 2-21-1 Osawa, Mitaka, Tokyo 181-8588, Japan}

\author[0000-0003-3618-7535]{Teruyuki Hirano}
\affiliation{Astrobiology Center, 2-21-1 Osawa, Mitaka, Tokyo 181-8588, Japan}
\affiliation{National Astronomical Observatory of Japan, 2-21-1 Osawa, Mitaka, Tokyo 181-8588, Japan}
\affiliation{Department of Astronomical Science, The Graduate University for Advanced Studies, SOKENDAI, 2-21-1 Osawa, Mitaka, Tokyo 181-8588, Japan}

\author[0000-0002-4677-9182]{Masayuki Kuzuhara}
\affiliation{Astrobiology Center, 2-21-1 Osawa, Mitaka, Tokyo 181-8588, Japan}
\affiliation{National Astronomical Observatory of Japan, 2-21-1 Osawa, Mitaka, Tokyo 181-8588, Japan}

\author[0000-0002-5051-6027]{Masashi Omiya}
\affiliation{Astrobiology Center, 2-21-1 Osawa, Mitaka, Tokyo 181-8588, Japan}
\affiliation{National Astronomical Observatory of Japan, 2-21-1 Osawa, Mitaka, Tokyo 181-8588, Japan}

\author[0000-0001-9194-1268]{Nobuhiko Kusakabe}
\affiliation{Astrobiology Center, 2-21-1 Osawa, Mitaka, Tokyo 181-8588, Japan}
\affiliation{National Astronomical Observatory of Japan, 2-21-1 Osawa, Mitaka, Tokyo 181-8588, Japan}



\begin{abstract}

Terrestrial exoplanets orbiting M-dwarf stars are promising targets for transmission spectroscopy with existing or near-future instrumentation. The atmospheric composition of such rocky planets remains an open question, especially given the high X-ray and ultraviolet flux from their host M dwarfs that can drive atmospheric escape. The 1.3 $R_\earth$ exoplanet GJ 486b ($T_{\rm{eq}} \sim$ 700\,K), orbiting an M3.5 star, is expected to have one of the strongest transmission spectroscopy signals among known terrestrial exoplanets. We observed three transits of GJ 486b using three different high-resolution spectrographs: IRD on Subaru, IGRINS on Gemini-South, and SPIRou on the Canada-France-Hawai'i Telescope. We searched for atmospheric absorption from a wide variety of molecular species via the cross-correlation method, but did not detect any robust atmospheric signals. Nevertheless, our observations are sufficiently sensitive to rule out several clear atmospheric scenarios via injection and recovery tests, and extend comparative exoplanetology into the terrestrial regime. Our results suggest that GJ 486b does not possess a clear H$_2$/He-dominated atmosphere, nor a clear 100\% water-vapor atmosphere. Other secondary atmospheres with high mean molecular weights or H$_2$/He-dominated atmospheres with clouds remain possible. Our findings provide further evidence suggesting that terrestrial planets orbiting M-dwarf stars may experience significant atmospheric loss.

\end{abstract}

\keywords{Exoplanet atmospheres (487) --- Planetary atmospheres (1244) 	--- Exoplanets (498) --- Exoplanet atmospheric composition (2021)}


\section{Introduction} \label{sec:intro}

Exoplanets that transit M-dwarf stars produce relatively deep transits due to their high planet-star radius ratios. Consequently, it may be possible to detect biosignatures in the atmospheres of terrestrial exoplanets with existing or near-future instrumentation \citep[e.g.,][]{Snellen2013, Wunderlich2019}. However, the long-term habitability of such planets depends crucially on the stability of their atmospheres, which are subject to high X-ray + extreme UV (XUV) emissions during the pre-main-sequence phase of their M-dwarf hosts \citep[e.g.,][]{Shkolnik2014, Peacock2020} as well as flare events \citep[e.g.,][]{NevesRibeirodoAmaral2022} that can drive atmospheric escape \citep[e.g.,][]{Vidal-Madjar2003, Bourrier2017b, Airapetian2017}. 

Terrestrial exoplanets are expected to exhibit a rich diversity in atmospheric compositions. Exoplanets with $R_{\rm{p}} < 1.4 \, R_\earth$ and orbital periods $<$ 100\,days likely form as gas-rich sub-Neptunes that subsequently lose their primordial atmospheres due to photoevaporation \citep{Rogers2021}. The amount of primordial atmosphere retained depends on several factors, such as their host star's spectral energy distribution and activity levels, along with how much gas they accrete during formation \citep[e.g.,][]{Bolmont2017, Owen2020}. Should the primordial hydrogen-dominated atmosphere be lost, a secondary atmosphere may be retained. The composition of secondary atmospheres depends on a variety of processes, including interior outgassing, impact delivery, erosion, atmosphere-surface exchange, weathering, and volatile sequestration \citep[e.g.,][]{Rogers2011}. Driven by these processes, hot terrestrial planets ($T \gtrsim 1500$ K) are predicted to have atmospheres containing atoms and ions, such as Na or Ca$^+$, produced by surface vaporization \citep{Schaefer2009} --- though other hydrogen- or nitrogen-rich compositions are possible \citep{Miguel2019}. However, cooler terrestrial planets are likely rich in molecular species, such as H$_2$O, CO$_2$, CO, O$_2$, C$_2$H$_2$, and CH$_4$ \citep[e.g.,][]{Ramirez2014, Madhusudhan2016, Kite2021}.

Given the range of theoretical possibilities, observational constraints on the composition of terrestrial exoplanet atmospheres are crucially needed. While there have been several attempts to detect the atmospheres of a few terrestrial exoplanets, these have only yielded constraints from nondetections.  For example, 55 Cancri e is a 1.9 $R_\earth$ exoplanet with a temperature of approximately 2400 K due to its 17 hr orbit around its G8V  host star \citep{Morris2021}.  Hubble Space Telescope observations suggest that 55 Cancri e may possess a thick atmosphere \citep{Tsiaras2016}. However, phase curve observations with the Spitzer space telescope \citep{Demory2016} and ground-based high-resolution transmission spectroscopy point toward 55 Cancri e not possessing a thick atmosphere or having no atmosphere at all \citep{Ridden-Harper2016, Esteves2017, Jindal2020, Deibert2021a, Zhang2021}.

A similar result was found for the cooler 1.3 $R_\earth$ exoplanet LHS 3844 b, which has an equilibrium temperature of 800 K and orbits an M5 host star \citep{Vanderspek2019}.  \citet{Kreidberg2019} observed the phase curve of  LHS 3844 b with the Spitzer space telescope and found that thick atmospheres with surface pressures greater than 10 bar could be ruled out, while less massive atmospheres would be susceptible to erosion by the stellar wind. Additionally, they found that a bare rock model fit their data well.  

Another candidate for atmospheric characterization that orbits an M-dwarf is GJ 1132b, which has a radius, mass, and equilibrium temperature of 1.1 $R_\earth$, 1.7 $M_\earth$, and 580 K, respectively \citep{Berta-Thompson2015}.  Using transit observations from the Hubble Space Telescope (HST), \citet{Swain2021} inferred that GJ 1132b possesses a H/He-rich atmosphere with large-amplitude spectral features. However, using the same HST data, \citet{Libby-Roberts2022} and \citet{Mugnai2021} independently found a featureless transmission spectrum, possibly indicating an atmosphere with a high mean molecular weight (MMW), a high-altitude aerosol layer, or a near-total lack of an atmosphere. These contrasting results highlight the difficulty of characterizing the atmospheres of small planets, even with space-based instrumentation.

Comprehensive searches for the atmospheres of the seven approximately Earth-sized planets that transit the M8V star TRAPPIST-1 have also been carried out. These transit observations with the Hubble and Spitzer space telescopes all point toward the absence of extended hydrogen-dominated atmospheres. However, denser, potentially habitable atmospheres are consistent with the observational limits for some TRAPPIST-1 planets \citep{deWit2016, Bourrier2017a, Bourrier2017b, deWit2018, Zhang2018, Wakeford2019, Gressier2022}.

GJ 486b\footnote{Alternative names include Wolf 437b and TOI 1827b.} is a 1.3 $R_\earth$ and 2.8 $M_\earth$ exoplanet with an equilibrium temperature of about 700 K, an M3.5 V host star of J-band magnitude 7.2, and an orbital period of 1.5 days \citep{Trifonov2021}.  These system properties make GJ 486b well suited to atmospheric characterization with transmission spectroscopy, as reflected by it having one of the largest transmission spectroscopy metrics \citep[as defined by][]{Kempton2018} of all known terrestrial planets \citep{Trifonov2021}.  Therefore, GJ 486b is an ideal observational target to inform our understanding of terrestrial planets around M-dwarf stars.

Previous studies have surmised that a wide range of atmospheric compositions are possible for GJ 486b. \cite{Caballero2022} refined GJ 486b's mass and radius, and presented a suite of interior and atmospheric models. They showed that possible atmospheres include: (i) a H/He-dominated atmosphere with solar abundances; (ii) a H/He atmosphere with enhanced metallicity (e.g., 100x solar abundances); (iii) a H$_2$O-dominated atmosphere; (iv) a CO$_2$-dominated atmosphere; or (v) bare rock with a tenuous atmosphere, possibly containing Na. Geochemical models based on initial conditions representative of a range of possible planets suggest that a H$_2$O-dominated atmosphere is the most likely outcome for a 3 $M_\earth$ planet like GJ 486b \citep{Ortenzi2020}. However, carbonaceous chondrite outgassing measurements suggest that CO$_2$-dominated atmospheres could also evolve on terrestrial planets \citep{Thompson2021}. Differentiating between these scenarios requires spectroscopic observations.

Molecular species such as those expected to be in GJ 486b's atmosphere produce dense forests of spectral lines that can be individually resolved with high-resolution spectroscopy.  The signals from these lines can be combined, even if they are individually hidden in the noise, by cross-correlating the data with a template containing the molecular lines \citep[e.g.,][]{Snellen2010, Birkby2018}.  This powerful technique has facilitated the detection of species such as H$_2$O, CO, HCN, CH$_4$, NH$_3$, and C$_2$H$_2$ in the transmission spectra of hot Jupiters \citep[e.g.,][]{ Giacobbe2021}.  As high-resolution spectroscopy probes line cores, which contain information about the upper (low-pressure) regions of exoplanet atmospheres, it is highly complementary to low-resolution observations from telescopes such as the JWST, which probe higher pressures \citep{Brogi2019}.  JWST is scheduled to observe two transits and eclipses of GJ 486b in programs GO 1981\footnote{\url{https://www.stsci.edu/jwst/science-execution/program-information.html?id=1981}} (P.I.: K. Stevenson) and GO 1743\footnote{\url{https://www.stsci.edu/jwst/science-execution/program-information.html?id=1743}} (P. I.: M. Mansfield), respectively.

Here, we present the first high-resolution transmission spectroscopy observations of GJ 486b. We use three transits to constrain GJ 486b's atmospheric composition. In what follows, in Sections~\ref{sec:data} and \ref{sec:DataReduction} we describe our observations and data reduction approach. Section~\ref{sec:model_spectra} describes our model spectra, while Section~\ref{sec:CrossCorrelationMethod} describes our cross-correlation methodology. Section~\ref{sec:Results} presents our results, while Section~\ref{sec:JWST} discusses the implications of our results for the upcoming JWST observations. Finally, Section~\ref{sec:conclusion} offers concluding remarks.

\section{Observations} \label{sec:data}

We used three different instruments to observe three separate transit events of GJ 486b. The first transit was observed with the Infrared Doppler (IRD) instrument on the 8.2 m Subaru telescope, providing a wavelength coverage of 0.97-1.75 $\mu$m at a resolving power of $R$ = 70k \citep{Tamura2012,Kotani2018}.  The second transit was observed with the Immersion GRating INfrared Spectrometer (IGRINS) on the 8.1 m Gemini-South telescope (program ID: GS-2021A-DD-105), providing a wavelength coverage of 1.45-2.45 $\mu$m at a resolving power of $R$ = 45k \citep{Park2014, Lee2016, Mace2018}.  The third transit was observed with the SPectropolarimètre InfraROUge (SPIRou) on the 3.6 m Canada-France-Hawai’i Telescope (CFHT) in queue mode (run ID: 21BC34), providing a wavelength coverage of 0.98-2.35 $\mu$m at a resolving power of $R$ = 75k \citep{Artigau2014}.  The observation start time, duration, integration time, cadence, total number of frames, number of frames in transit, number of frames out of transit, and average signal-to-noise (S/N) ratio per pixel for these observations are shown in Table \ref{tab:ObsInfo}.

\begin{table*}[!htb]
\caption{\label{tab:ObsInfo} Summary of Observations}
\begin{tabular}{ccccccccc} \hline  \hline
Instrument & Ob. Date & Ob. Start & Duration & Ob. Phase & Exp.  & Cadence & No. Frames & Avg. S/N \\ 
& (UT) & ~~ time (UT) &  (hr) & range  &  time (s) & (s) & (in/out) & per pixel \\ \hline
IRD & 2021-04-18 & 07:52:27  & 2.40 & $-$0.0355 - 0.0332 & 200 & 211 & 41 (12/29) & 130 \\
IGRINS  & 2021-05-25 & 00:18:52 & 2.36 & $-$0.0286 - 0.0364 & 10 & 64 & 133 (57/76)$^*$ & 130 \\
SPIRou & 2022-01-23 & 13:07:05 & 2.33 & $-$0.0363 - 0.0298 & 184 & 200 & 42 (17/25)$^\dagger$ & 150 \\
\hline
\multicolumn{9}{l}{{$^*$} For IGRINS, the first frame (which was out of transit) and the 78th frame (which was in transit) were excluded} \\
\multicolumn{9}{l}{~~~  from the analysis because their S/Ns were about 70\% of the average S/N per frame.}\\

\multicolumn{9}{l}{{$^\dagger$} The 31st SPIRou frame (which was in transit) was excluded from the analysis because its S/N was about 60\% } \\
\multicolumn{9}{l}{~~~of the average S/N per frame.} \\

\end{tabular}

\end{table*}

\section{Data Reduction and Processing} \label{sec:DataReduction}

The IRD transit data were reduced following the approach of \cite{Hirano2020}, which uses a custom code to correct for a known count bias that depends on the readout channel \citep{Kuzuhara2018}. Subsequent reduction steps were performed using the \texttt{echelle} package of \texttt{IRAF}. As these observations did not use the laser frequency comb, the wavelength calibration was based on the Th-Ar lamp.  The wavelength solution was stable to within approximately 0.04 km s$^{-1}$ over the course of the observations. This stability was determined by first cross-correlating every frame with a model telluric spectrum produced using SkyCalc \citep{Noll2012, Jones2013} to find the telluric radial velocity of each frame. We then calculated the standard deviation of all telluric radial velocities during the night (after applying a barycentric correction). The model produced by SkyCalc is highly customizable, but the most important parameters are target elevation or airmass and the precipitable water vapor.   

The IGRINS transit data were reduced by the IGRINS Pipeline Package \citep[PLP;][]{Sim2014,Mace2018}. Though this pipeline provides wavelength-calibrated data, the PLP wavelength calibration drifted by about 1 kms$^{-1}$ through the course of the observations. To fit and correct this drift, 
we first cross-correlated SkyCalc telluric models (as above) with every frame in spectral orders that had no saturated telluric lines.  The spectral orders used for these cross-correlations are shown in Table \ref{tab:ExcludedWavesForIGRINS}.  For each frame, we then summed the cross-correlation functions (CCFs) from all our considered orders to produce a master CCF.  Next, we fit a Gaussian to the master CCF of each frame and adopted the Gaussian center as the radial velocity of the observed telluric spectrum. Finally, we fit the telluric radial velocities with a third-degree polynomial as a function of time and used this to shift the data onto a common wavelength grid using linear interpolation. The amplitudes of the corrective shifts were approximately 0.5$-$1 km s$^{-1}$.

The SPIRou transit data were reduced by the observatory using the SPIRou Data Reduction Software (DRS) \footnote{\url{http://www.cfht.hawaii.edu/Instruments/SPIRou/SPIRou_pipeline.php}}.  This pipeline provides data with and without a telluric correction, and we opted to use the telluric-corrected data. 

For all data sets, after the above procedures, we normalized the spectra on a per-order basis to have a continuum flux of 1. We then masked all spectral regions with a normalized flux less than 0.3. For the IGRINS data, we also masked telluric emission lines, as in \citet{Rousselot2000} and \cite{Oliva2015}. 

We then applied the \texttt{SYSREM} algorithm \citep{Tamuz2005} to all data sets to remove remaining telluric features and other systematic trends.  We applied \texttt{SYSREM} to our data in fluxes (without first converting to magnitudes), as in \citet{Gibson2020} and \citet{Nugroho2021}. Because \texttt{SYSREM} incorporates uncertainties, we estimated the uncertainties of the normalized data as 
\begin{equation}
\sigma = \frac{\sigma_t \otimes \sigma_\lambda}{\sigma_{t,\lambda}}
\end{equation}

where $\sigma_t$ and $\sigma_\lambda$ are the standard deviations in the time and wavelength dimensions, respectively, $\sigma_{t,\lambda}$ is the standard deviation in both the time and wavelength dimensions, and $\otimes$ is the outer product operator. 

To determine an effective number of \texttt{SYSREM} iterations to use, we followed a well-established approach and examined how each order's variance decreased as a function of the number of applied \texttt{SYSREM} iterations  \citep[e.g.,][]{Deibert2021a, Herman2022}. For each order, we examined plots of the variance as a function of the number of \texttt{SYSREM} iterations and identified by eye the number of iterations where the decrease in data variance began to plateau. We noted the number of \texttt{SYSREM} iterations determined for each order and then used the most common number of iterations as a constant for each data set.  This resulted in us applying five \texttt{SYSREM} iterations to all orders of the SPIRou and IRD data and 10 \texttt{SYSREM} iterations to all orders of the IGRINS data. While our chosen values are likely not the formal optimal values, they are a close approximation, as telluric features and other systematic trends were removed effectively. In general, previous studies have found that the exact number of \texttt{SYSREM} iterations has little to no effect on the final results \citep[e.g.,][]{Deibert2021b}. Some IGRINS orders, as indicated in Table \ref{tab:ExcludedWavesForIGRINS}, had near-total telluric extinction across most of their spectral range. Such severe contamination cannot be corrected by \texttt{SYSREM}, so we excluded these orders from the analysis. Finally, for all data sets, we removed any outlying points more than three standard deviations away from the mean.

Barycentric corrections for the IRD and IGRINS data were calculated with the \texttt{radial\_velocity\_correction} function in \texttt{Astropy}. The same corrections for the SPIRou data were provided by the DRS.

\section{Model Spectra} \label{sec:model_spectra}

We searched for chemical species in GJ 486b's atmosphere through a cross-correlation analysis with model transmission spectra. We generated a grid of template spectra using the open-source radiative transfer code \texttt{petitRADTRANS}\footnote{https://petitradtrans.readthedocs.io/en/latest/} \citep{Molliere2019b}. Each template contains one or more absorbing chemical species embedded in a background of spectrally inactive gas. Specifically, we searched for a wide range of gaseous absorbers: C$_2$H$_2$, CH$_4$, CO, CO$_2$, FeH, H$_2$O, HCN, H$_2$S, K, Na, NH$_3$, PH$_3$, SiO, TiO, and VO. We used the default opacity data included with \texttt{petitRADTRANS} \citep[see][]{Molliere2019b} for all species. However, for CH$_4$ we used both the default opacity data from Exomol \citep{Yurchenko2014} and our custom-made opacity grid based on a line list adopted from HITEMP 2020 \citep{Hargreaves2020}. We produced the custom opacity grid following the pressure-temperature grid in the petitRADTRANS website (e.g., PTgrid.dat). We calculated the cross section using \texttt{HELIOS-K} 2.0 \citep[]{Grimm2015, Grimm2021} with a line wing cutoff of 100 cm$^{-1}$ at a resolution of 0.001 cm$^{-1}$ which was later resampled at a constant resolution of $R = 10^6$. Our templates include two chemical models: (i) single chemical species models and (ii) multispecies chemical equilibrium models.

Our single-species models span a range of volume mixing ratios (VMRs) and MMWs. The VMRs range from $\log_{10}$(VMR) = $-$7 to 0 (assumed constant with altitude). We varied the MMWs independently of the VMRs, subject to the following boundary conditions: (i) the minimum MMW corresponds to a H$_2$-dominated atmosphere with the template species (i.e. $\rm{MMW}_{\rm{min}} \approx 2 \cdot (1 - \rm{VMR}_i) + \rm{VMR}_i \cdot \rm{MMW}_i$); and (ii) the maximum MMW corresponds to an N$_2$-dominated atmosphere with the template species (i.e. $\rm{MMW}_{\rm{max}} \approx 28 \cdot (1 - \rm{VMR}_i) + \rm{VMR}_i \cdot \rm{MMW}_i$). These boundary conditions ensure the MMWs are physically consistent with the presence of a spectrally inactive gas mixture with unknown ratios of H$_2$, He, or N$_2$. These models assumed an isothermal pressure-temperature (P-T) profile at GJ 486b's equilibrium temperature: 700 K \citep{Trifonov2021}. 

\begin{figure*}[!htb]
\centering
\includegraphics[width=\textwidth]{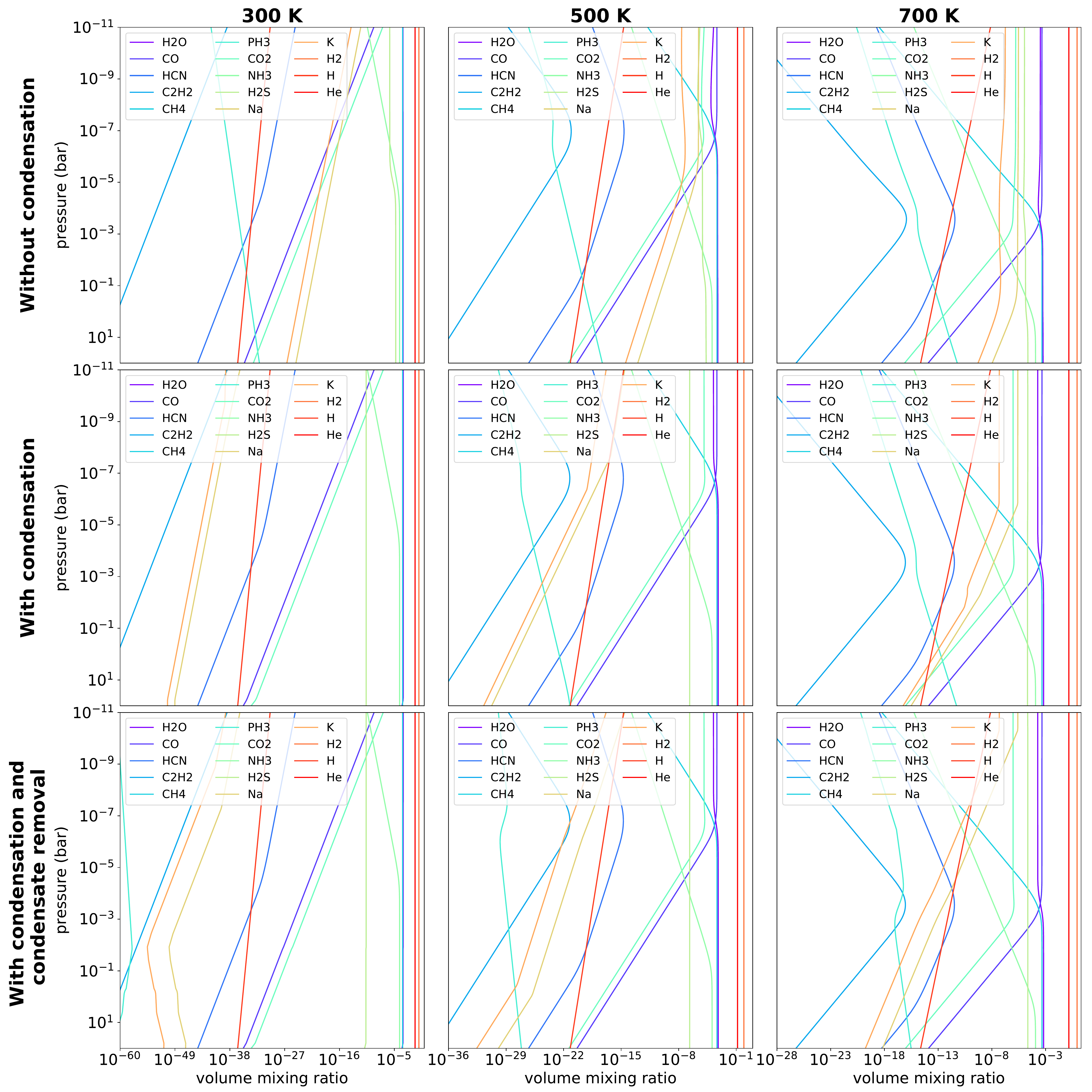} 
\caption{Chemical equilibrium abundance profiles for GJ 486b computed by GGchem. Columns: isothermal pressure-temperature profiles at 300 K (left), 500 K (middle), and 700 K (right). Rows: no condensation (top), condensation included (middle), condensation and rain out (bottom).}
\label{fig:GGchem_abundances}
\end{figure*}

Our second set of models included all the above mentioned spectrally active species with thermochemical equilibrium abundances. We computed equilibrium abundances for a solar composition atmosphere using the open-source code \texttt{GGchem}\footnote{\url{https://github.com/pw31/GGchem}} \citep{Woitke2018}, assuming isothermal P-T profiles at 400, 500, 600, and 700 K. We considered three different approaches for modeling condensation in GGchem: (1) no condensation, (2) condensation included, and (3) condensation included, but condensates removed (e.g., rain out). These abundance profiles are shown in Fig.~\ref{fig:GGchem_abundances}.  Although condensation would likely result in the production of aerosols such as clouds or hazes, we do not attempt to account for the effect of such aerosols on the resulting transmission spectrum.

Our \texttt{petitRADTRANS} models were computed line by line at $R = 10^6$ using 130 layers spaced evenly in log-pressure from $10^{-11}$ to $10^2$ bar. We then resampled the model spectra to the resolution of the data ($R$ = 45 k, 70 k, and 75 k for IGRINS, IRD, and SPIRou, respectively) using the \texttt{SpectRes} package\footnote{\url{https://spectres.readthedocs.io/en/latest/}} \citep{Carnall2017} which preserves the integrated flux. Because cross-correlations are not sensitive to continua, we subtracted the continuum from our model spectra (after binning to the data resolution) to produce our final cross-correlation templates.

\section{Cross-correlation Method} \label{sec:CrossCorrelationMethod}

To search for GJ 486b's atmosphere, we cross-correlated the models described in Section \ref{sec:model_spectra} with the processed data from Section \ref{sec:DataReduction}.

We cross-correlated each model with one processed frame at a time, to produce a cross-correlation function (CCF). We phase-folded these CCFs by shifting them into the planet rest frame. GJ 486b's radial velocity as a function of time, relative to its host star, is given by

\begin{equation}
v_r = K_\mathrm{P} \sin(2 \pi \phi(t))    
\label{eqn:rv}
\end{equation}

where $K_\mathrm{P} = v_{orb} \, \sin(i)$, $v_{orb}$ is GJ 486b's orbital velocity, $i$ is its orbital inclination, and $\phi(t)$ is the orbital phase at the time, $t$, of the observations. We calculated $\phi(t)$ assuming a circular orbit and the ephemeris reported by \cite{Trifonov2021}.  At this stage, we also accounted for the Earth's barycentric radial velocity with the corrections described in Section \ref{sec:DataReduction}.

To optimally combine our three data sets, we mapped our cross-correlation functions to log-likelihoods \citep[following, e.g.,][]{Brogi2019,Gibson2020,Herman2022}. This approach allows the log-likelihoods from each data set to be trivially summed.

The log-likelihood for each data set, $\ln \mathcal{L}_{D}$, is calculated according to 

\begin{equation}
\label{eqn:lnL}
    \ln \mathcal{L}_D = -\frac{N}{2} \mathrm{ln} \frac{\chi^2}{N}
\end{equation}

where the $D$ subscript refers to the data set (e.g., IRD, IGRINS, and SPIRou), $N$ is the total number of in-transit data points in the data set, and $\chi^2$ is related to the cross-correlation function, $\mathrm{CCF}$, via

\begin{equation} \label{eqn:chi2}
    \chi^2_D = \sum \frac{f_i^2}{\sigma_i^2} + \alpha^2 \sum \frac{m_i^2}{\sigma_i^2} - 2\alpha ~\mathrm{CCF}
\end{equation}

where $f_i$ is the mean-subtracted spectrum, $\sigma_i$ is the outer product of the standard deviation of each wavelength and exposure bin (normalized by the standard deviation of the spectra in each order), $m_i$ is the mean-subtracted Doppler-shifted model, $\alpha$ is a scaling factor to allow uncertainty in the scale of the model, and the $\mathrm{CCF}$ is defined as
\begin{equation}
    \textrm{CCF}(v_\mathrm{sys}) = \sum_i \frac{f_i m_i(v_\mathrm{sys})}{\sigma_i^2}.
\end{equation}
We determined which frames were in transit by generating model transit light curves with the BAsic Transit Model cAlculatioN in Python (\texttt{BATMAN}\footnote{\url{https://github.com/lkreidberg/batman}}) package \citep{Kreidberg2015}, using the system parameters reported by \citet{Trifonov2021}, and weighted all in-transit frames equally. We converted the summed log-likelihood to a normalized likelihood, $\mathcal{L}_\mathrm{norm}$, according to  $\mathcal{L}_\mathrm{norm} = \exp(\mathrm{ln} \mathcal{L} - \mathrm{max}(\mathrm{ln} \mathcal{L}))$.

\section{Results}  \label{sec:Results}

\subsection{Cross-correlation search} \label{sec:CrossCorrelationSearch}

\begin{figure*}[!htb]
\centering
\includegraphics[width=\textwidth]{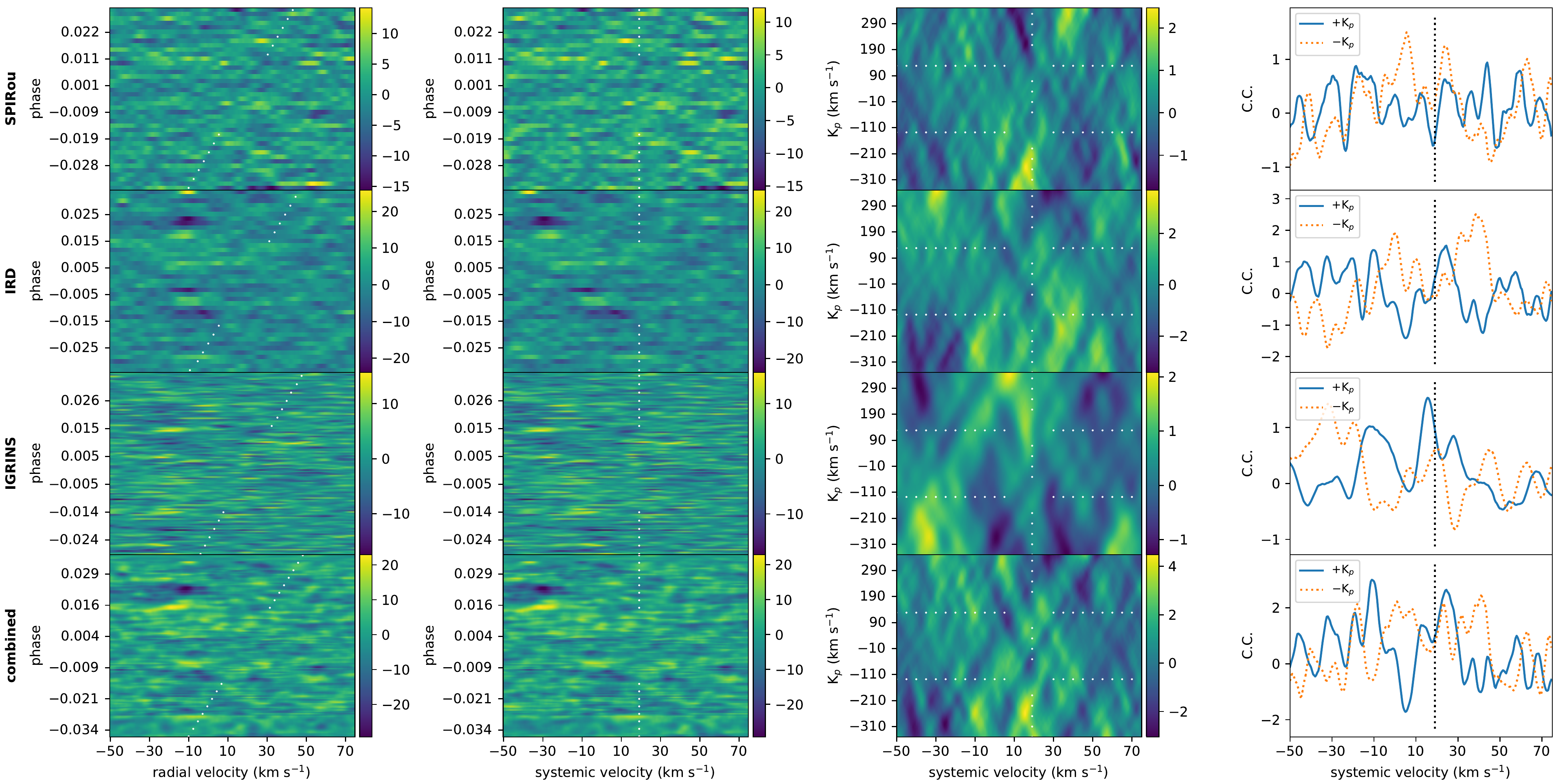} 
\caption{The results of cross-correlating the SPIRou (first row), IRD (second row), and IGRINS (third row) data sets with a model transmission spectrum containing spectral lines only from H$_2$O. The bottom row shows the result of combining all data sets together. The first column shows the cross-correlation for each frame in the barycentric frame. The expected trace of the planet's signal is shown as the dotted diagonal line.  The second column shows the cross-correlation for each frame in the planet's rest frame. The dotted line shows the expected trace of the planet's signal at the system's systemic velocity. The third column shows the phase-folded cross-correlation signal as a function of phase. The dotted horizontal and vertical lines show the planet's expected position in $\pm$$K_\mathrm{P}$ and systemic velocity, respectively. The fourth column shows the 1D cross-correlation as a function of systemic velocity at $\pm$$K_\mathrm{P}$. The black dotted line indicates the system's systemic velocity.}
\label{fig:CCF_H2O}
\end{figure*}

\begin{figure*}[!htb]
\centering
\includegraphics[width=\textwidth]{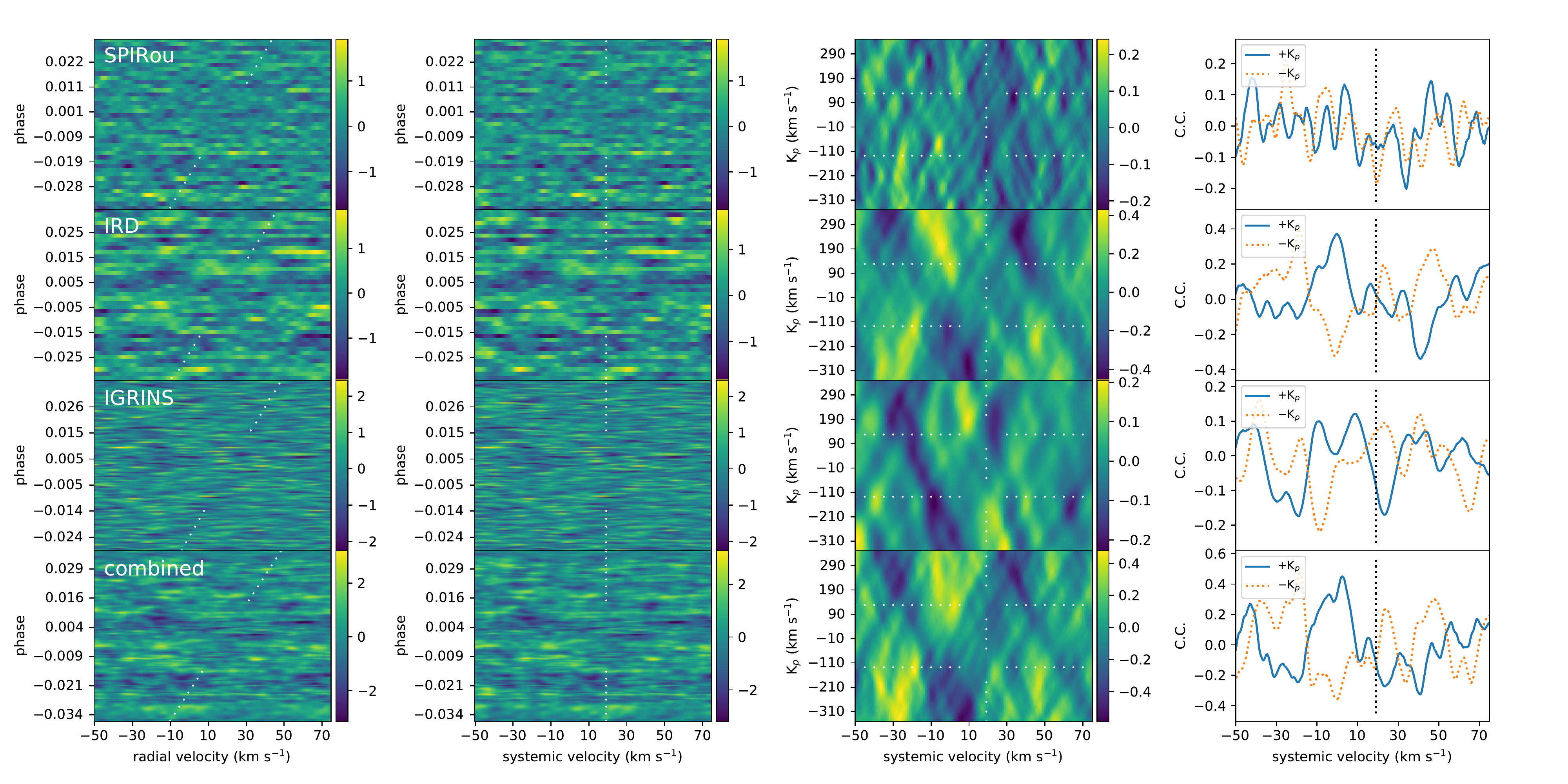} 
\caption{Same as Fig. \ref{fig:CCF_H2O} except for CO$_2$.}
\label{fig:CCF_CO2}
\end{figure*}

\begin{figure*}[!htb]
\centering
\includegraphics[width=\textwidth]{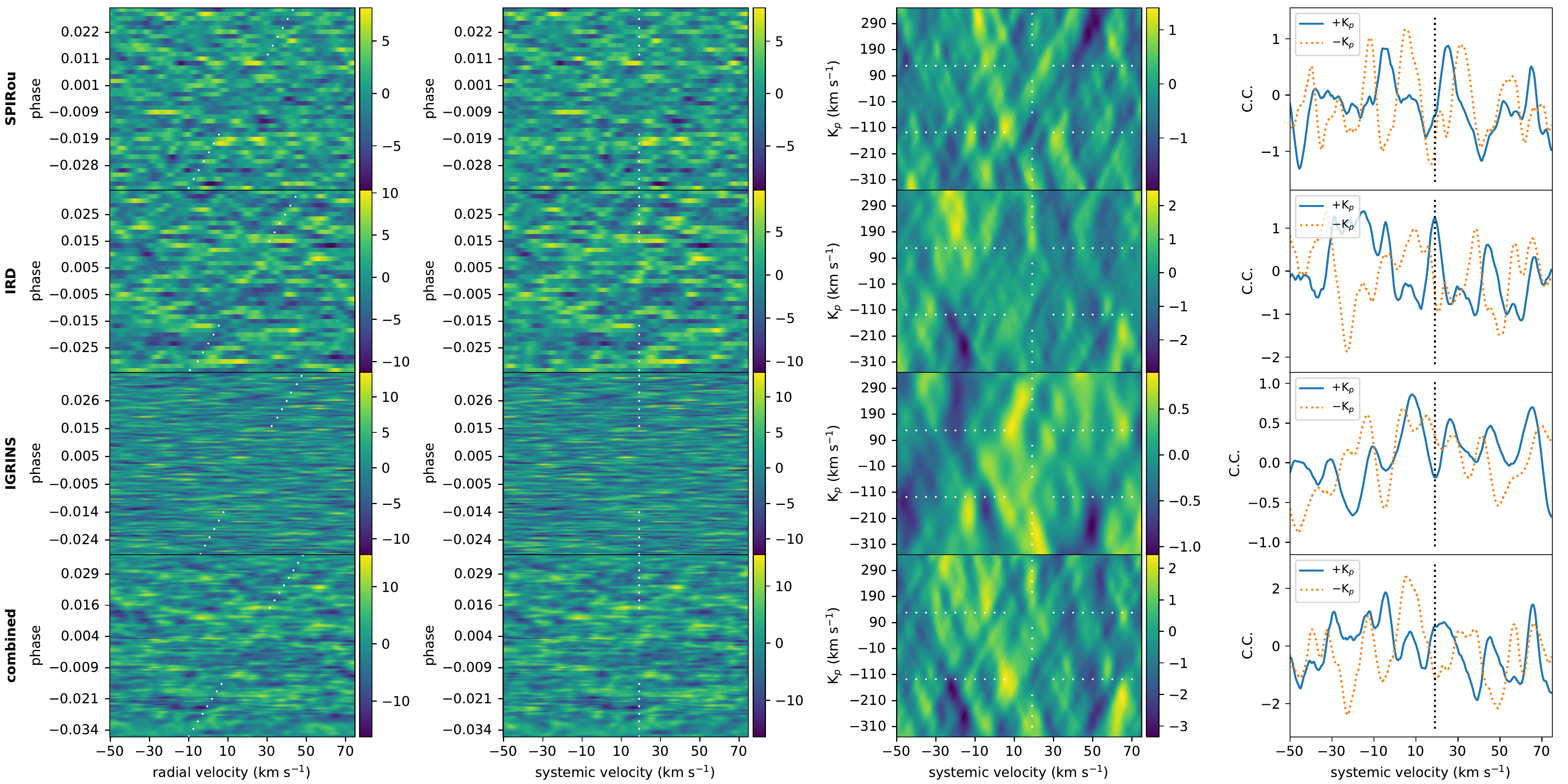} 
\caption{Same as Fig. \ref{fig:CCF_H2O} except for HCN.}
\label{fig:CCF_HCN}
\end{figure*}

\begin{figure*}[!htb]
\centering
\includegraphics[width=\textwidth]{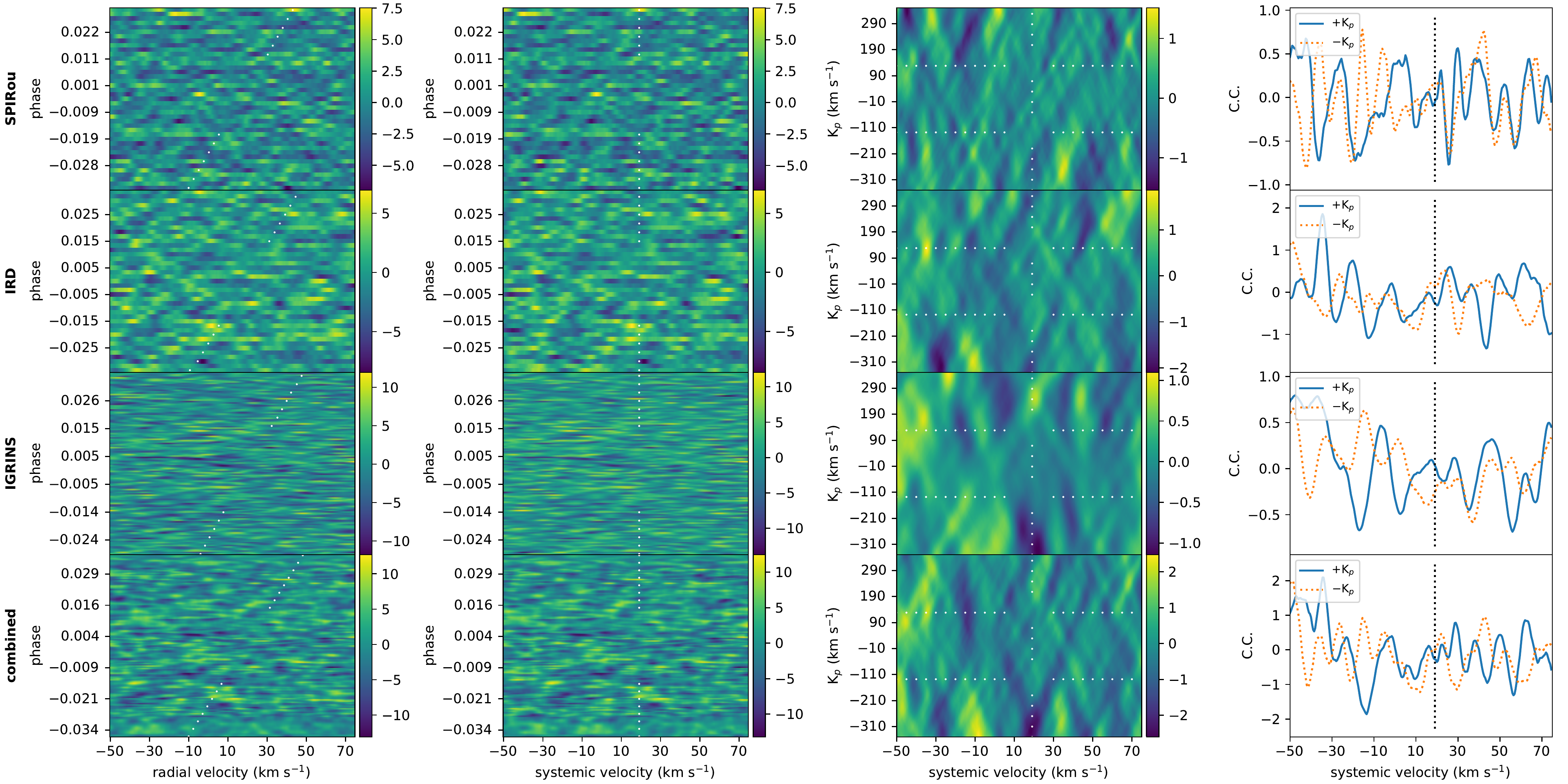} 
\caption{Same as Fig. \ref{fig:CCF_H2O} except for NH$_3$.}
\label{fig:CCF_NH3}
\end{figure*}

\begin{figure*}[!htb]
\centering
\includegraphics[width=\textwidth]{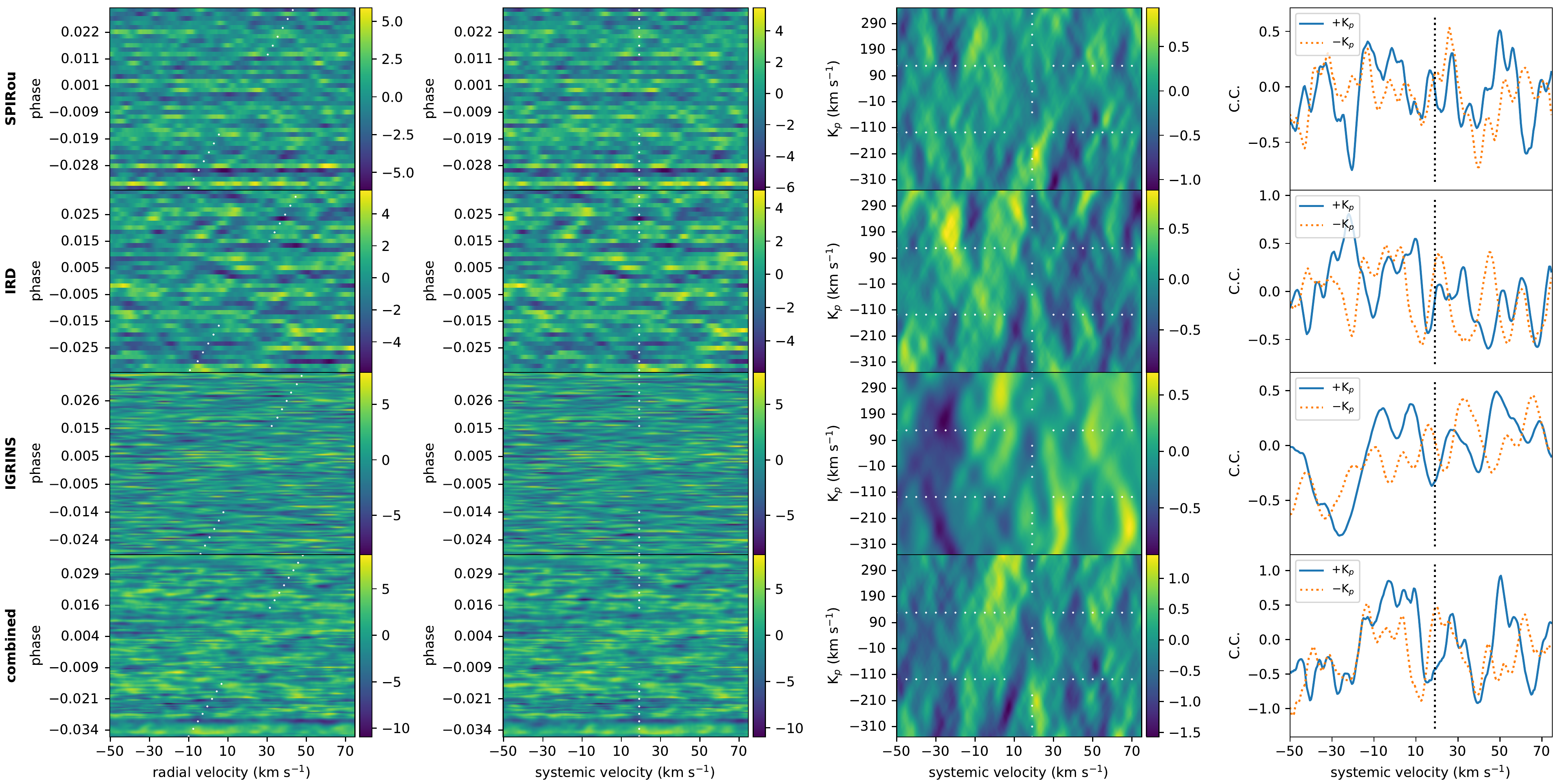} 
\caption{Same as Fig. \ref{fig:CCF_H2O} except for CH$_4$.}
\label{fig:CCF_CH4}
\end{figure*}

\begin{figure*}[!htb]
\centering
\includegraphics[width=\textwidth]{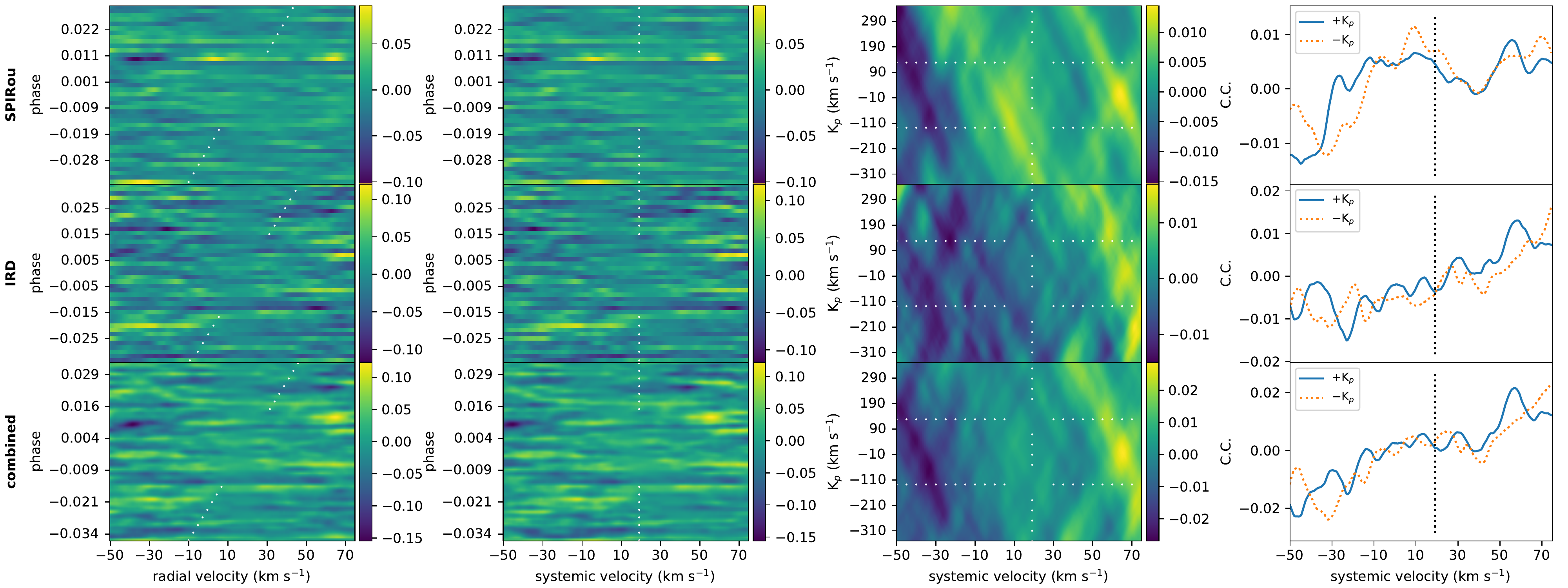} 
\caption{Same as Fig. \ref{fig:CCF_H2O} except for Na. The row for IGRINS is omitted because no Na lines are within its spectral range.}
\label{fig:CCF_Na}
\end{figure*}

\begin{figure*}[!htb]
\centering
\includegraphics[width=\textwidth]{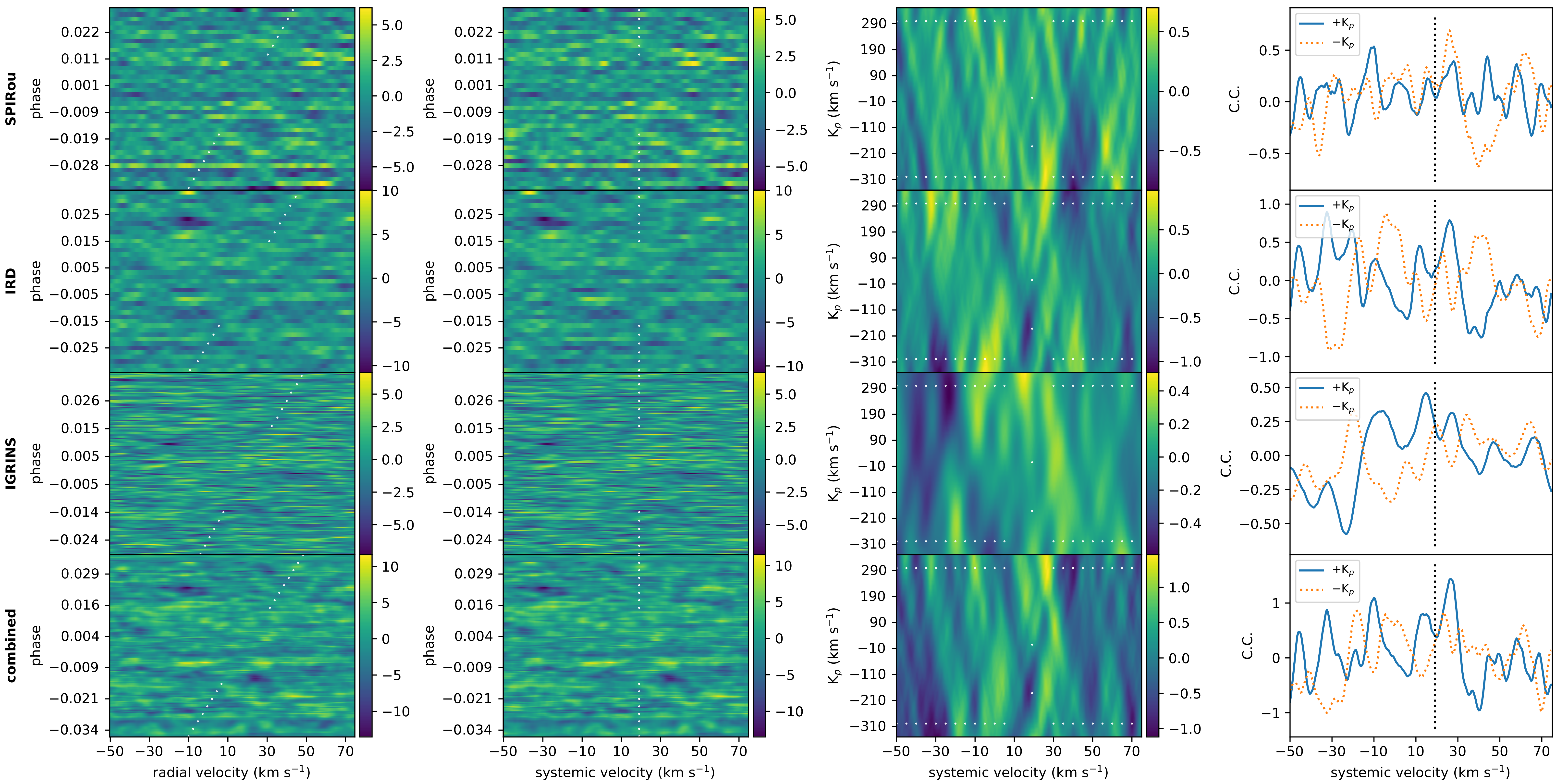} 
\caption{Same as Fig. \ref{fig:CCF_H2O} except for using the solar abundances at 700 K model generated by GGchem.}
\label{fig:CCF_ggchem700K}
\end{figure*}

We cross-correlated the models described in Section \ref{sec:model_spectra} with the processed data and searched for peaks in the phase-folded cross-correlation function at the expected $K_\mathrm{P}$ and $v_\mathrm{sys}$ of GJ 486b.  Figs. \ref{fig:CCF_H2O} $-$ \ref{fig:CCF_ggchem700K} show the results of our search for H$_2$O, CO$_2$, HCN, NH$_3$, CH$_4$, Na, and the 700 K solar abundance model with condensation, respectively.  These figures show the 1D cross-correlation functions phase-folded using GJ 486b's expected value of $K_\mathrm{P}$ as well as its negative value. We show the result of using $-K_\mathrm{P}$ as we later use it to inject model spectra into the data in Section \ref{sec:InjectionAndRecoveryTests}.

\begin{figure*}[!htb]
\centering
\includegraphics[width=0.6\textwidth]{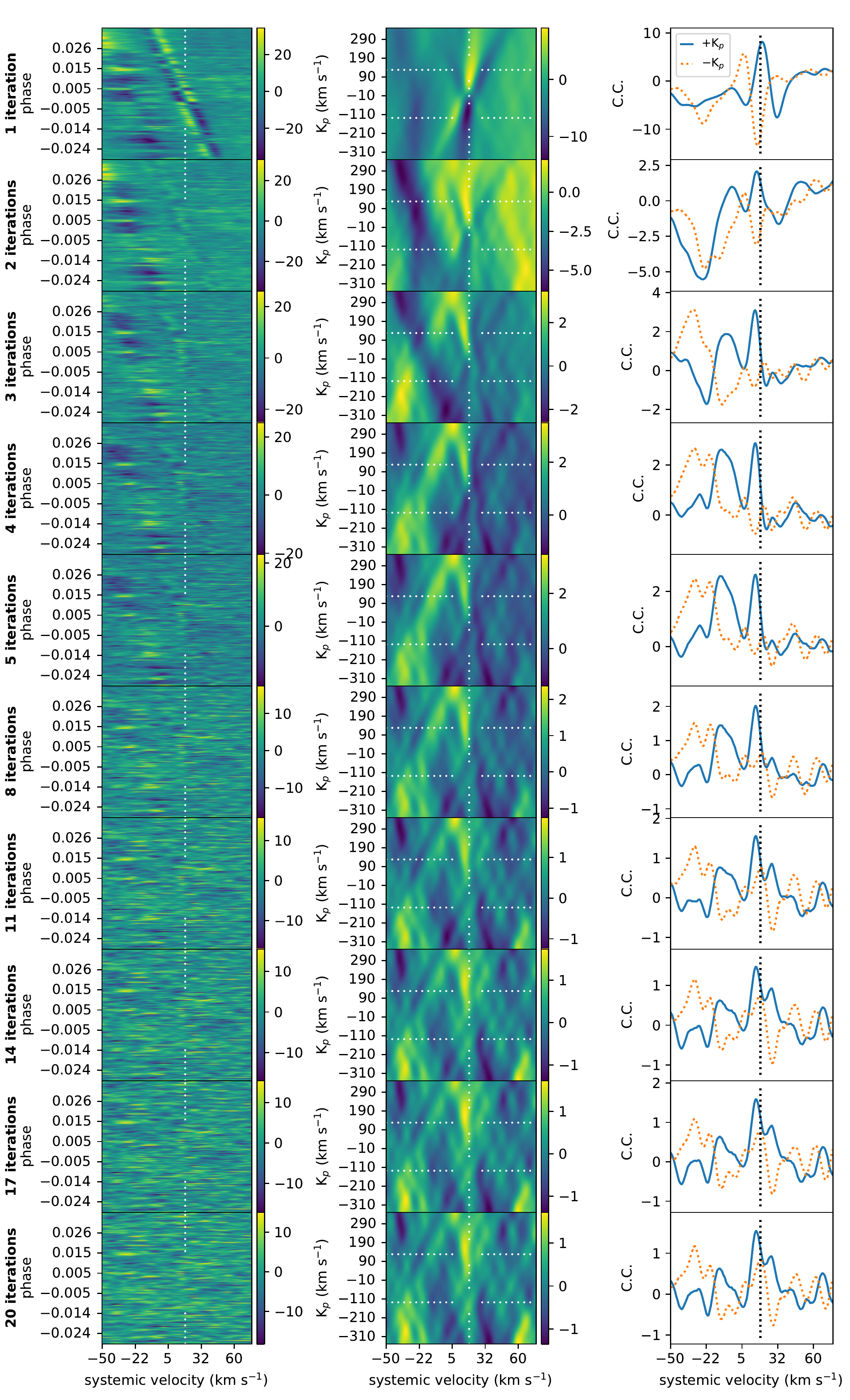} 
\caption{The results of cross-correlating a model transmission spectrum containing spectral lines only from H$_2$O with the IGRINS data after a variable number of SYSREM iterations were applied.  The rows show the result of applying 1, 2, 3, 4, 5, 8, 11, 14, 17, and 20 iterations, respectively. The first column shows the cross-correlation for each frame in the planet's rest-frame. The dotted line shows the expected trace of the planet's signal at the system's systemic velocity. The second column shows the phase-folded cross-correlation signal as a function of phase. The dotted horizontal and vertical lines show the planet's expected position in $\pm$$K_\mathrm{P}$ and systemic velocity, respectively.  The thrid column shows the 1D cross-correlation as a function of systemic velocity at $\pm$$K_\mathrm{P}$. The black dotted line indicates the system's systemic velocity.}
\label{fig:IGRINS_water_CCF_SYSREM}
\end{figure*}

We find no evidence for robust correlations between any of our models\footnote{Including both the Exomol \citep{Yurchenko2014} and HITEMP 2020 \citep{Hargreaves2020} opacity data for CH$_4$.} and the data. The most promising feature is of H$_2$O in the IGRINS data set (see Fig. \ref{fig:CCF_H2O}), which is at the 3$\sigma$ level\footnote{Approximated as the amplitude of the cross-correlation function at the expected velocity divided by the standard deviation of the surrounding regions.} in the expected position in $K_\mathrm{P}$ and $v_\mathrm{sys}$.  However, it is likely that this feature is produced by imperfectly removed water lines in the spectrum of GJ 486b's M-dwarf host star, given that it is present after applying only one iteration of \texttt{SYSREM}, as shown in Fig. \ref{fig:IGRINS_water_CCF_SYSREM}. 

We do not understand why these residuals were not removed by \texttt{SYSREM} and only appear in the IGRINS data set. However, we note that they appear to be qualitatively similar to those reported by \cite{Brogi2013}.  In their case, \cite{Brogi2013} detected molecular absorption in the dayside emission spectrum of 51 Pegasi~b on two separate occasions, but not a third occasion when strong stellar residuals were present in the data.  \citet{Chiavassa2019} revisited this data and used 3D simulations of stellar convection to remove the stellar spectrum.  This method better suppressed the stellar residuals in the third observation and allowed the planetary signal to be detected.  In a future study, it may be beneficial to investigate whether the method of \citet{Chiavassa2019} can better suppress the stellar residuals in our IGRINS data.

Regarding our nondetection of Na, \citet{Caballero2022} suggest that if GJ 486b were to have a tenuous atmosphere with a surface pressure of 10$^{-6}$ bar, nonthermal processes such as surface sputtering by ions may heat its atmosphere to temperatures on the order of 10,000 K.  They show that a Na abundance of 10$^{-9}$ in such an atmosphere would produce transit depths of approximately 1\% in the cores of the Na D lines (5889.95 and 5895.92 \AA), which could be detected by observing only a few transits with the Calar Alto high-Resolution search for M dwarfs with Exoearths with Near-infrared and optical Echelle Spectrographs \citep[CARMENES;][]{Quirrenbach2014}.  The spectral ranges of our observations do not include the Na D lines, but our IRD and SPIRou data sets do include the Na lines at 1.138 and 1.141 $\mu$m, which are much weaker and more difficult to detect.  Therefore, while we found no evidence for a nonthermal Na atmosphere, we do not consider this to be evidence for its absence. We recommend further investigation of this scenario with additional observations targeting the Na D lines.

\subsection{Constraining GJ 486b's Atmospheric Composition \label{sec:InjectionAndRecoveryTests}}

As we did not detect the atmosphere of GJ 486b, here we determine the abundances that our data can rule out under the assumption that GJ 486b's atmosphere is clear. We do this by injecting cloud-free model spectra into the reduced data before processing and cross-correlating as in Sections \ref{sec:DataReduction} and \ref{sec:CrossCorrelationMethod}.  For each in-transit frame, we injected the model transmission spectrum at the velocity given by Equation \ref{eqn:rv} if $K_\mathrm{P}$ were replaced by its negative value. We injected the signals at $-K_\mathrm{P}$ so that they would not be affected by any weak real planet signals that may be present in the data.  

\begin{figure*}[!htb]
\centering
\includegraphics[width=0.6\textwidth]{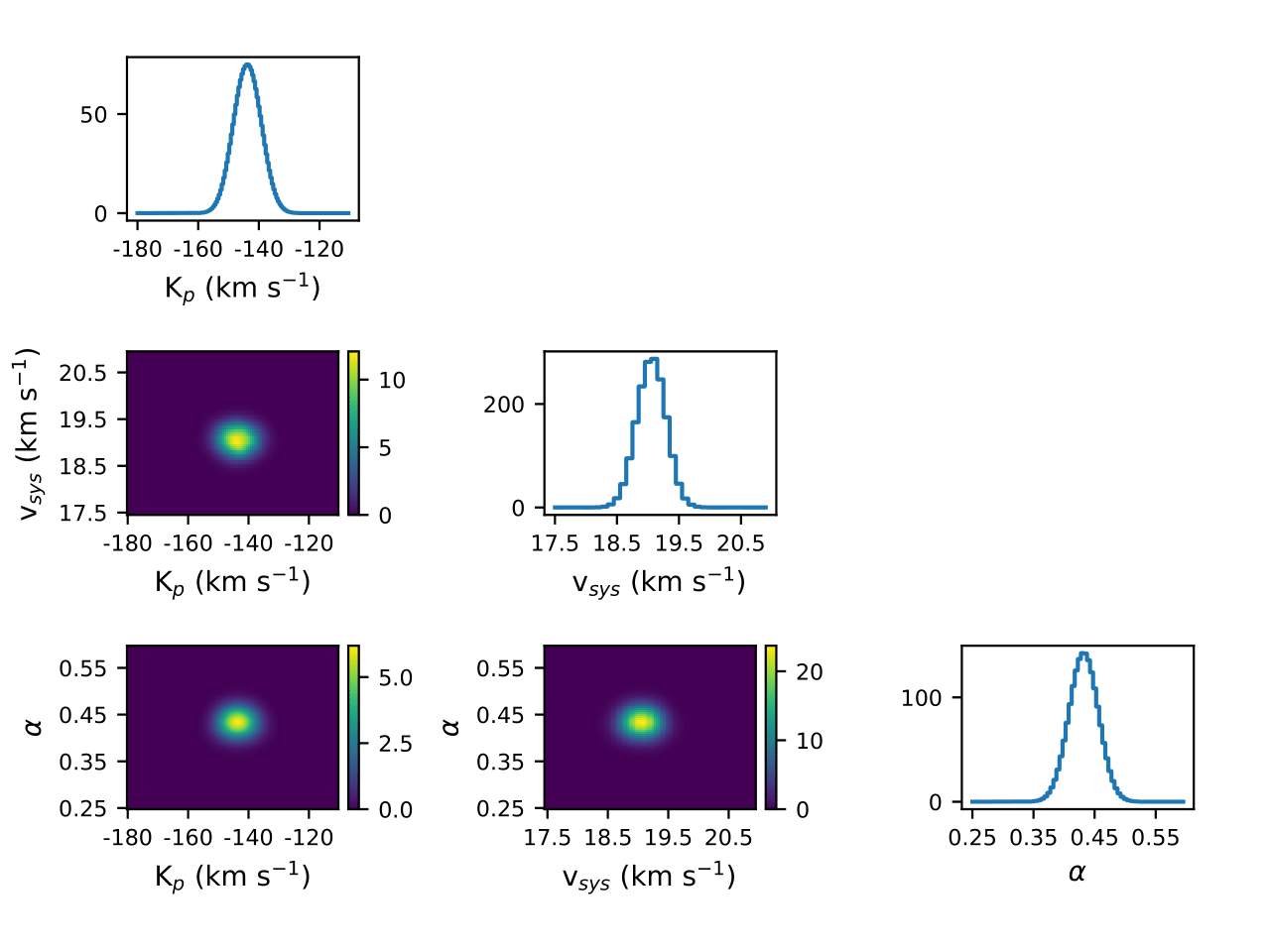} 
\caption{Example of marginalized likelihoods for a strong injected signal}
\label{fig:CornerPlotExample}
\end{figure*}

To optimally combine the different data sets, we summed the log-likelihoods.  Our cross-correlation to log-likelihood mapping produced an array with dimensions $K_\mathrm{P}$, $v_\mathrm{sys}$, and $\alpha$. Examples of the recovered conditional (2D) and marginalized (1D) likelihood distributions after injecting a water model into the data are shown in Fig. \ref{fig:CornerPlotExample}.

We used the marginalized distributions of $K_\mathrm{P}$, $v_\mathrm{sys}$, and $\alpha$ (as shown in Fig. \ref{fig:CornerPlotExample}) to find their most likely values and their associated 1$\sigma$ errors.  With this approach, the significance to which an injected signal was recovered is well approximated by how well $\alpha$ is constrained \citep{Gibson2020, Gibson2022}. This method works because we assume that marginalized distributions of $\alpha$ that are constant at $\alpha$ = 0, or unconstrained, correspond to a nondetection. Therefore, we are essentially measuring how significantly $\alpha$ differs from zero.

However, in practice, our recovered values of $\alpha$ were slightly less than the true values, leading to a slight underestimate of the true recovered detection significances.  This is because we did not explicitly account for how \texttt{SYSREM} distorts an underlying planet signal when removing the stellar and telluric lines.  \citet{Gibson2022} showed that this can be corrected by pre-processing the models in an analogous way to how \texttt{SYSREM} processes the planet signal in the data. We investigated the feasibility of using this approach by trying it for a single model spectrum and a typical grid covering $\pm$60 km s$^{-1}$ around the expected $K_\mathrm{P}$ and $v_\mathrm{sys}$ with step sizes of  1 km s$^{-1}$. The run time of this trial on a reasonably capable workstation\footnote{With the following specifications: Ryzen 5900X (12 core) CPU, 128 GB of DDR 3200 RAM, and a PCIe 4.0 M.2 SSD.} was 80 minutes. Our full analysis uses about 640 different model spectra, which would therefore require a total run time of about 36 days.  We consider this to be infeasible for this study, so we instead report our upper limits with the caveat that they likely underestimate the true upper limits. 

\begin{figure*}
\centering
\begin{tabular}{cc}
   H$_2$O & CH$_4$ \\
  \includegraphics[width=0.5\textwidth]{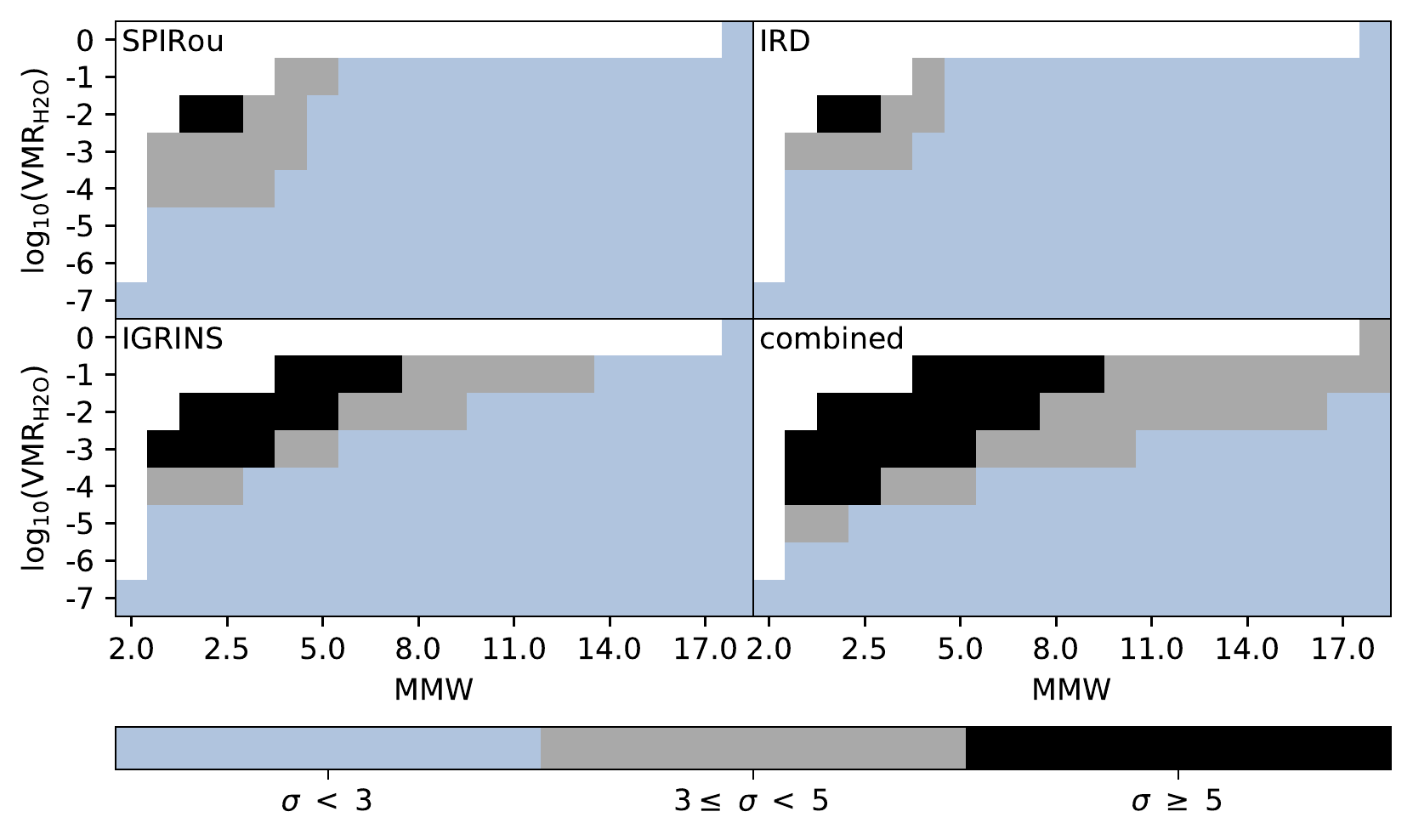} & \includegraphics[width=0.5\textwidth]{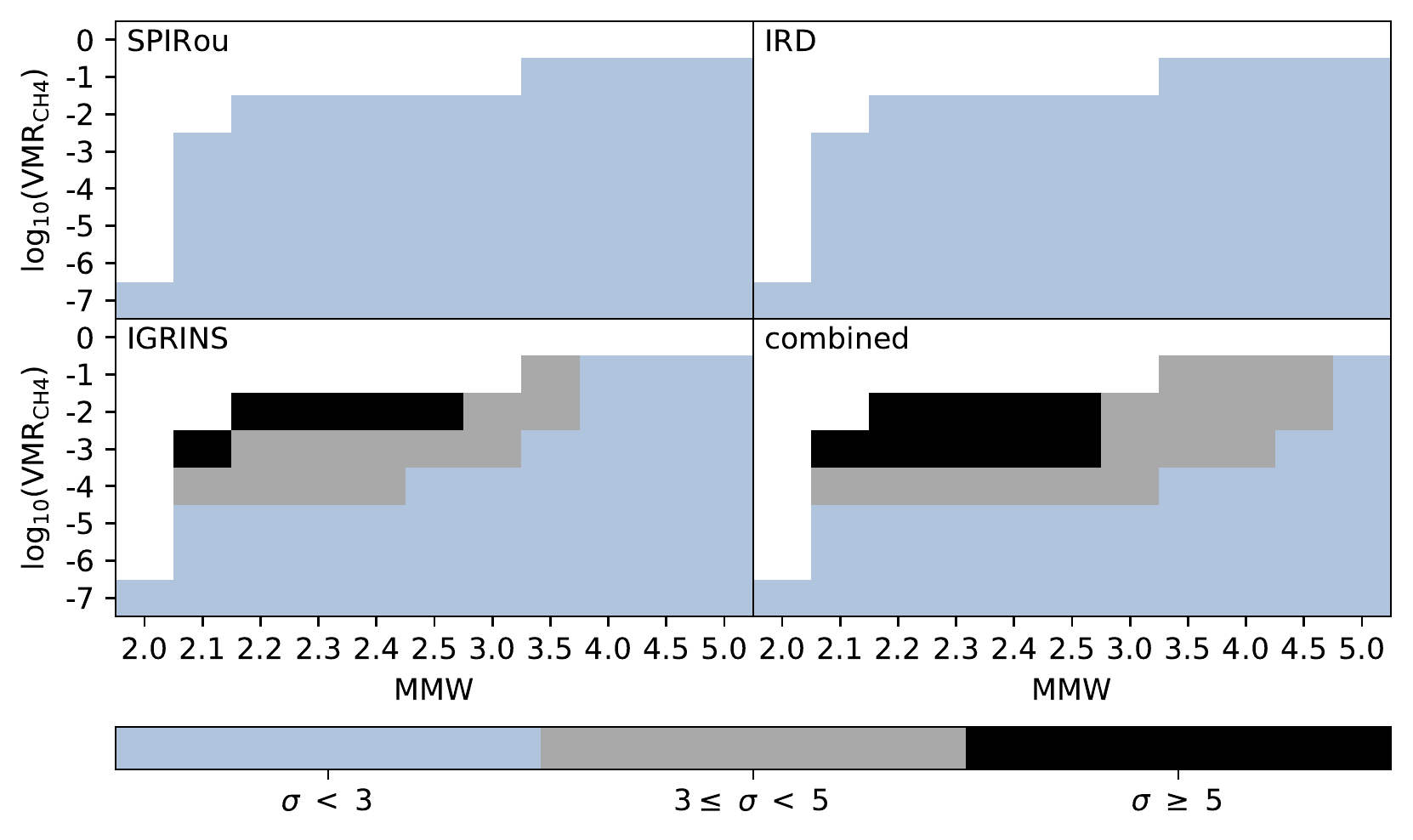} \\ 
     NH$_3$ & HCN \\
  \includegraphics[width=0.5\textwidth]{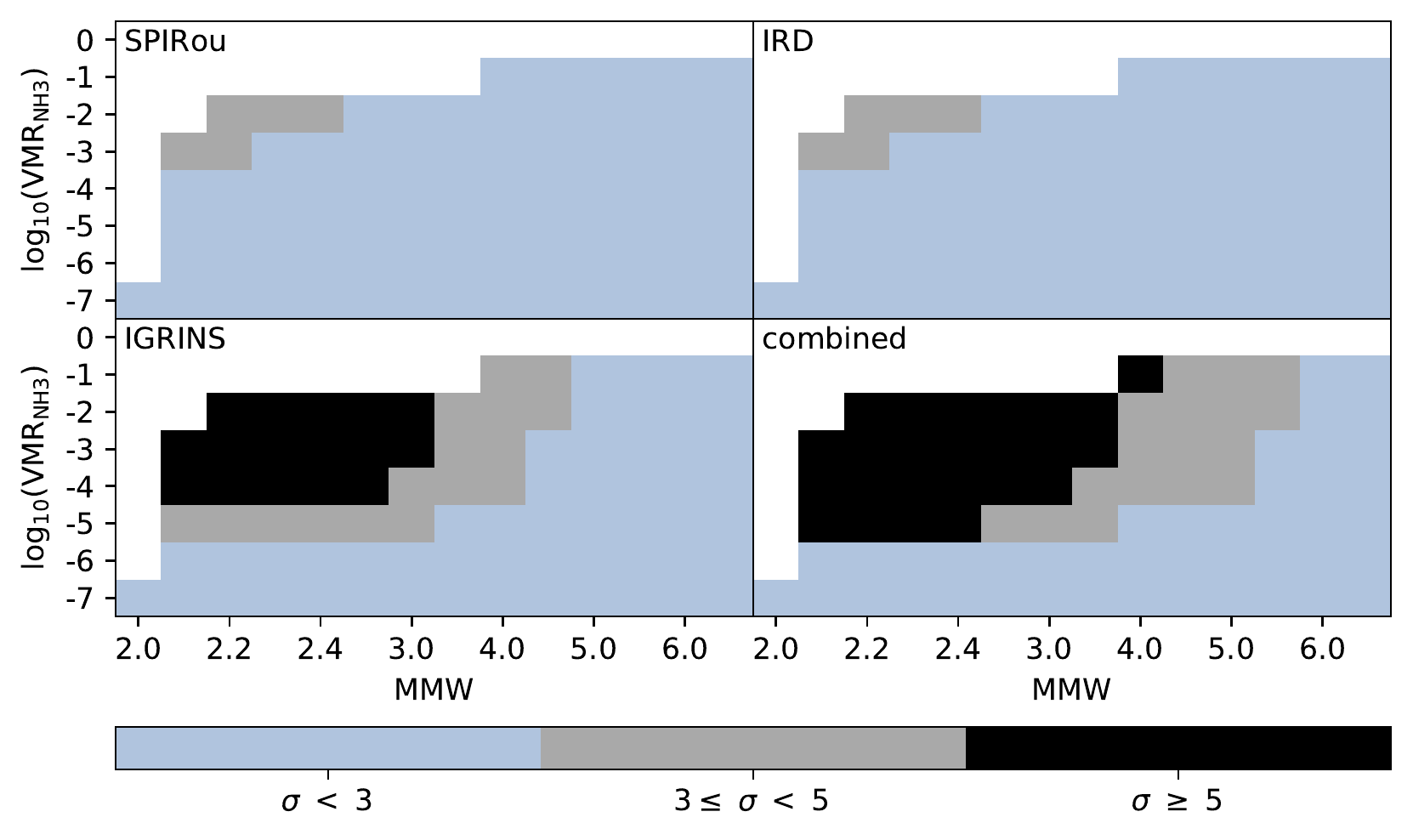} & \includegraphics[width=0.5\textwidth]{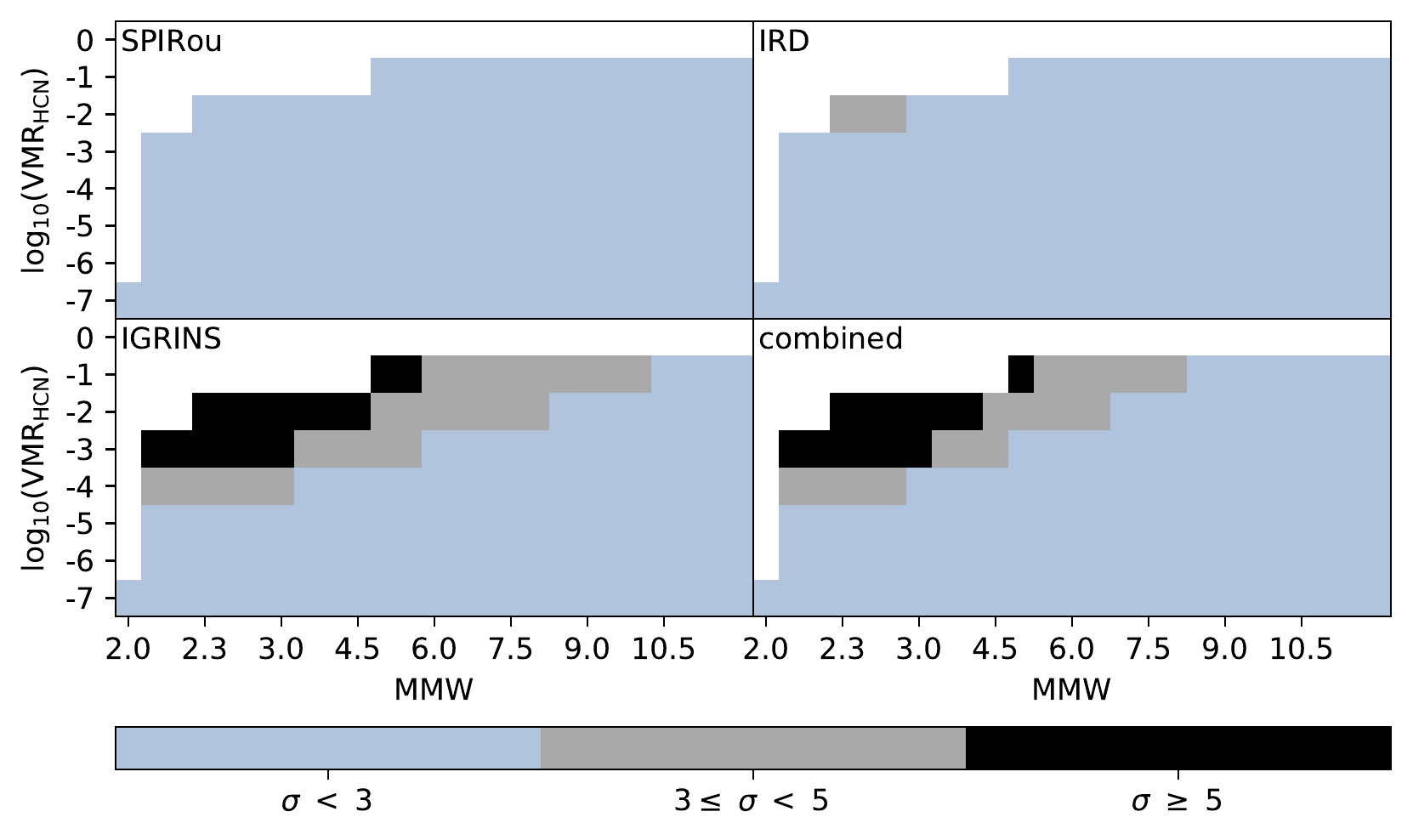} \\ 
     CO$_2$  & CO \\
  \includegraphics[width=0.5\textwidth]{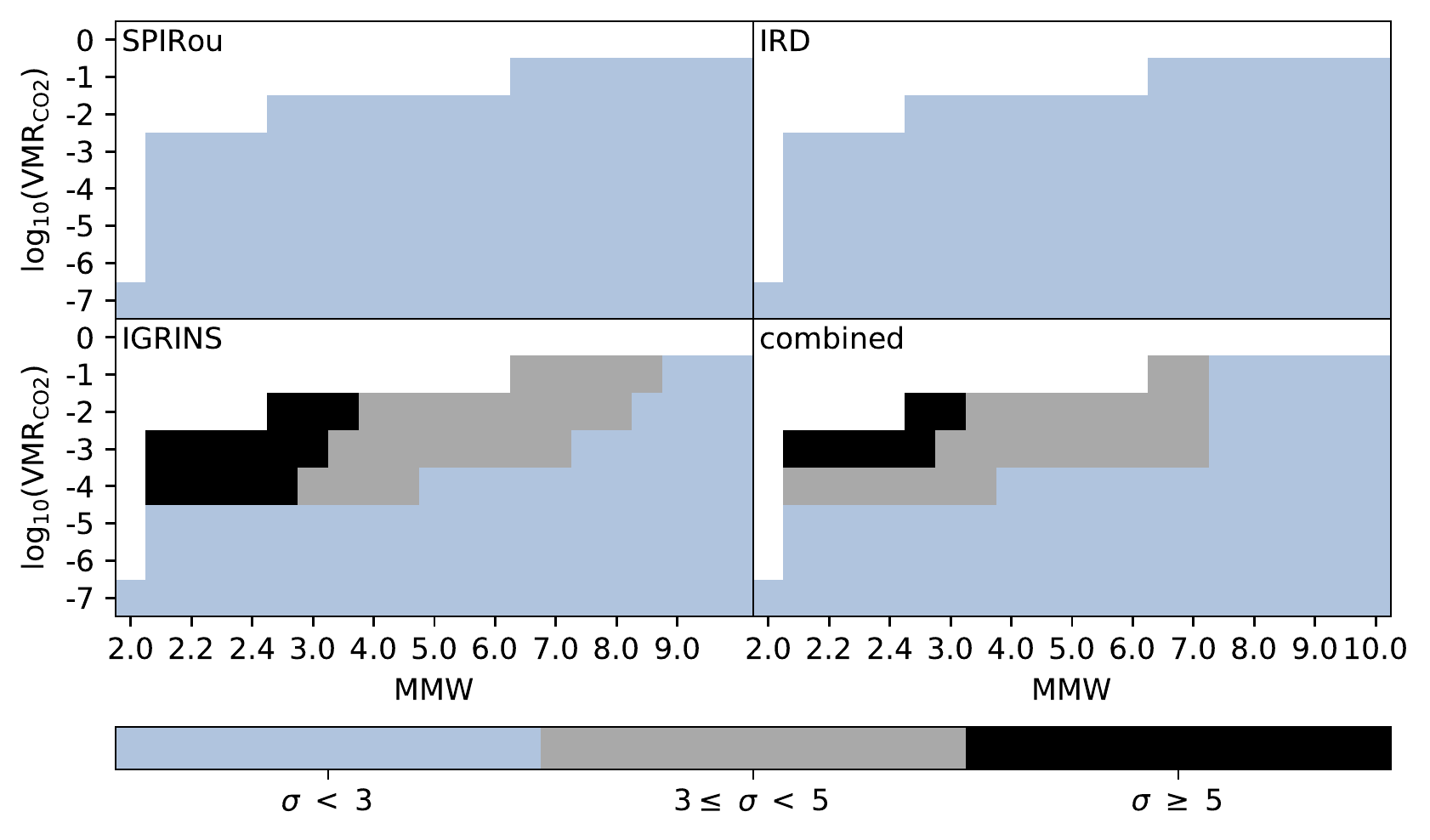} & \includegraphics[width=0.5\textwidth]{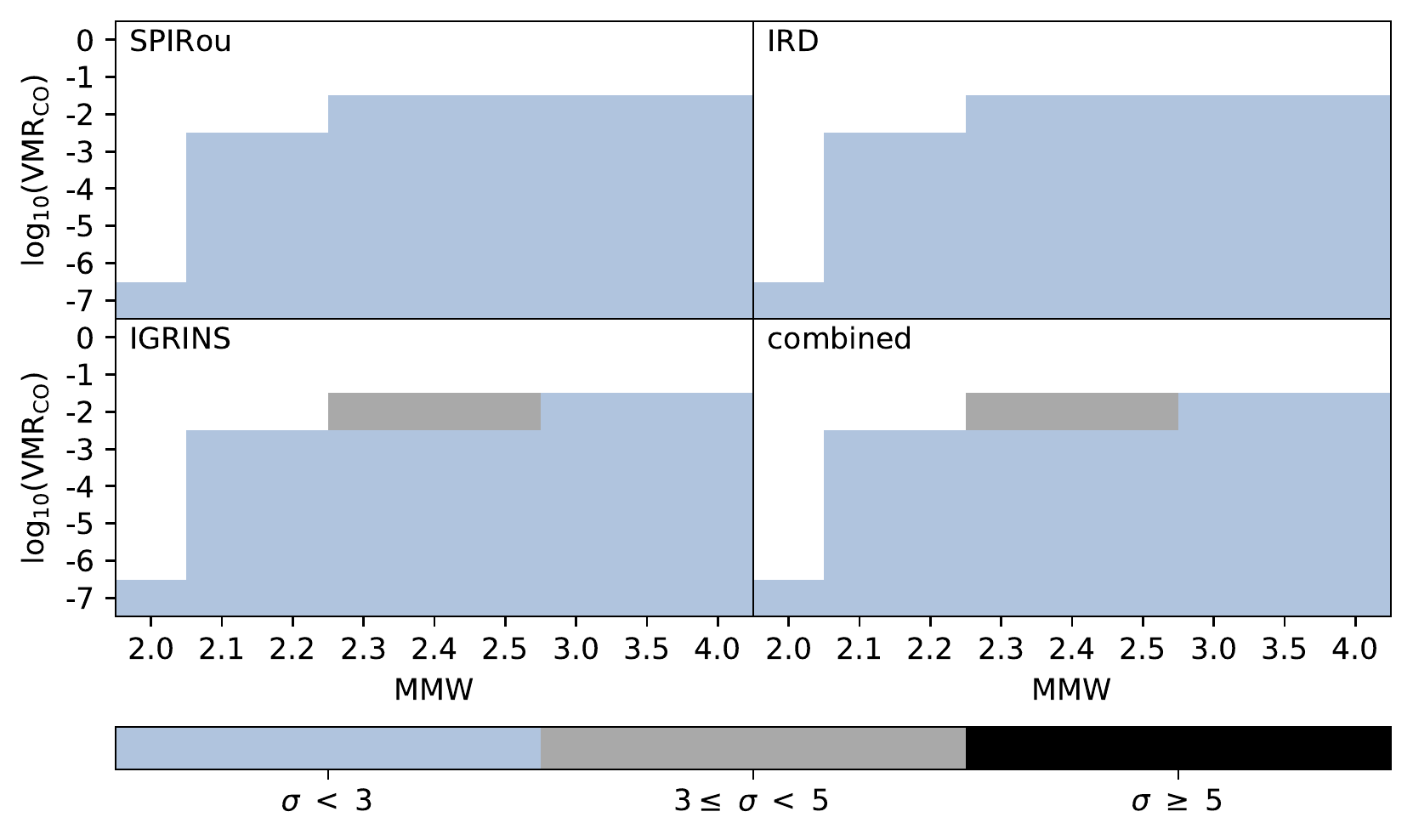} \\ 
  \end{tabular}
\caption{Constraints on the presence of H$_2$O (top left), CH$_4$ (top right), NH$_3$ (middle left), CH$_4$ (middle right), CO$_2$ (lower left), CO (lower right) in GJ 486b's atmosphere. Shown within each large panel are the limits from SPIRou (top left), IRD (top right), IGRINS (bottom left), and all data sets combined (lower right). The vertical axis shows the $\log_{10}$ of the VMR of the given species, while the horizontal axis shows the atmosphere's MMW. The black and gray regions indicate the VMR-MMW parameter space that can be ruled out to 5$\sigma$ and 3$\sigma$, respectively. The light blue regions are allowed by our observations.}
\label{fig:CombinedConstraints}
\end{figure*}

We were able to recover some injected models for H$_2$O, CH$_4$, NH$_3$, HCN, CO$_2$, and CO, indicating that our high-resolution observations can constrain the abundances of these chemical species under the assumption of GJ 486b having a clear atmosphere. As these species have different molecular weights, line strengths, and numbers of lines, they are each sensitive to different regions of VMR-MMW parameter space. These constraints are shown in Fig \ref{fig:CombinedConstraints} for H$_2$O, CH$_4$, NH$_3$, HCN, CO$_2$, and CO, respectively.  Each figure shows the VMR-MMW constraints for the individual data sets (SPIRou, IRD, IGRINS) and all data sets combined. Models with higher MMWs have smaller atmospheric scale heights, making them more difficult to recover. Injected models that were recovered to significances of $\geq$5$\sigma$ and 3$\leq$$\sigma$$<$5 are colored in black and gray, respectively. Injected models that were recovered to significances of 0$\leq$$\sigma$$<$3 are considered to have not been recovered and are shown in light blue.  This means that we can rule out model atmospheres to 5$\sigma$ and 3$\sigma$ in the VMR-MMW regions colored black and gray, respectively.  Conversely, the light blue regions represent atmospheric compositions that are allowed by our data.

We obtain strong constraints on the presence of H$_2$O in GJ 486b's atmosphere. Assuming a clear atmosphere, our observations rule out $\log_{10}$(VMR) $\geq$ $-$3 for atmospheric MMWs $\leq$ 5 and $\log_{10}$(VMR) $\geq$ $-$4 for MMWs $\leq$2.5 to 5$\sigma$. Therefore, our observations are inconsistent with a clear H$_2$/He-dominated atmosphere with solar (or somewhat subsolar) H$_2$O abundances. We can also, less confidently, rule out a clear atmosphere of $\log_{10}$(VMR) $\geq$ $-2$ for MMWs $\leq$ 16 and a clear 100\% H$_2$O atmosphere to 3$\sigma$. Our results, therefore, suggest that GJ 486b does not have a clear H$_2$O-dominated atmosphere. This could have important implications for planetary interior models, which often predict H$_2$O-dominated atmospheres as the most likely possibility for 3 $M_\earth$ planets like GJ 486b \citep[e.g.,][]{Ortenzi2020}.  

\begin{figure*}[!htb]
\centering
\includegraphics[width=0.6\textwidth]{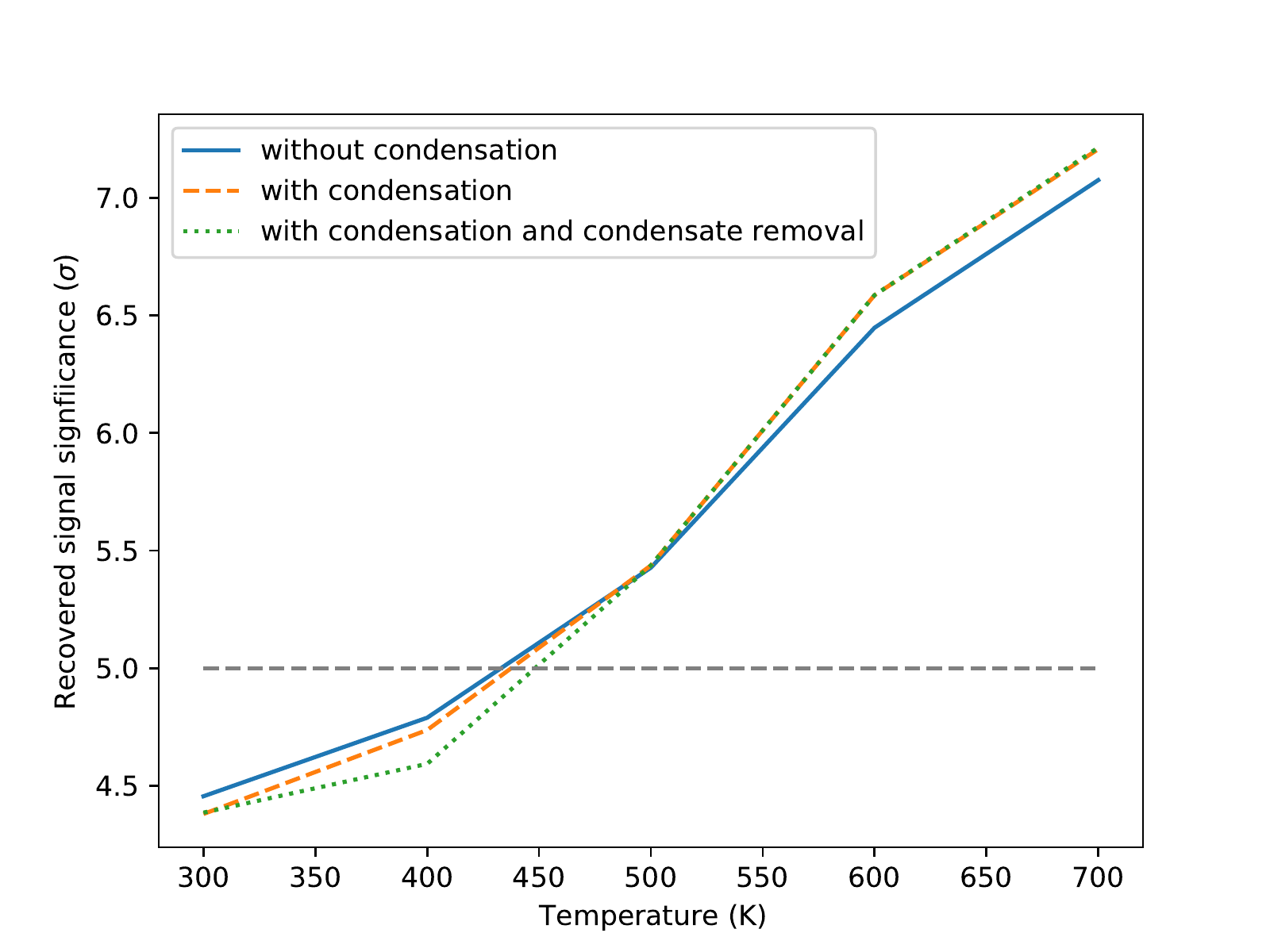} 
\caption{The significance to which injected cloud-free solar abundance models are recovered from our data as a function of temperature for GJ 486b. These models were generated by GGchem and used different approaches for modeling the effect of condensation on the abundance profiles.  However, the transmission spectra resulting from these models do not account for clouds or haze that may form from this condensation.  The blue solid line did not account for condensation, the orange dashed line did account for condensation, and the green dotted line accounted for condensation but condensates were removed to model rain out.}
\label{fig:temperature_constraints}
\end{figure*}

Our derived constraints on the abundances of CH$_4$, NH$_3$, HCN, CO$_2$, and CO are summarized in Table \ref{tab:AbundancesRuledOut}.  These constraints show that our observations are inconsistent with a clear H$_2$/He-dominated atmosphere with a solar CH$_4$ abundance. However, solar abundances of NH$_3$, HCN, CO$_2$, and CO are consistent with our data.  For cloud-free models including all the above species at solar abundances, we rule out to 5$\sigma$ those with temperatures $>$ 500 K, as shown in Fig. \ref{fig:temperature_constraints}. 

\begin{table*}[!htb]
\centering
\caption{\label{tab:AbundancesRuledOut} VMRs Ruled Out by Our Data for Given Atmospheric MMWs}
\begin{tabular}{ccccccc} \hline \hline
        & Approx. Solar   &\multicolumn{2}{c}{Ruled Out to 5$\sigma$} & ~~~~ & \multicolumn{2}{c}{Ruled Out to 3$\sigma$} \\ 
Species & $\log_{10}$(VMR) & $\log_{10}$(VMR) $\geq$ & MMW $\leq$      & ~~~~ &  $\log_{10}$(VMR) $\geq$ &  MMW $\leq$     \\ \hline
H$_2$O  & $-$3.3  & $-$3       &    5.0                     & ~~~~ &                        0 & 18              \\
        &   & $-$4       &    2.5                     & ~~~~ &                   $-$2   &  16             \\ 
CH$_4$  & $-$3.3  & $-$3       &    2.5                     & ~~~~ &                   $-$2   &  4.5            \\
        &   &  -         &    -                       & ~~~~ &                   $-$4   &  3.0            \\
NH$_3$  & $-$6.3  & $-$3       &    3.5                     & ~~~~ &                   $-$4   &  5.0            \\
        &   & $-$5       &    2.4                     & ~~~~ &                   $-$5   &  3.5             \\
HCN     & $-$13 & $-$2       &    4.0                     & ~~~~ &                   $-$2   &  6.5            \\
        &   & $-$3       &    3.0                     & ~~~~ &                   $-$4   &  2.5             \\
CO$_2$  & $-$8.3  & $-$2       &    3.0                     & ~~~~ &                   $-$3   &  7.0            \\
        &   & $-$3       &    2.5                     & ~~~~ &                   $-$4   &  3.5            \\
CO      & $-$6  & -          &    -                       & ~~~~ &                   $-$2   &  2.5            \\
\hline
\end{tabular}
\end{table*}

We were not able to recover any of our injected models for C$_2$H$_2$, FeH, H$_2$S, Na, K, PH$_3$, SiO, TiO, or VO. This indicates that, regardless of their VMR or the background atmosphere MMW, our observations are not sensitive to these chemical species. 

For all the species considered here, the data from IGRINS on Gemini-S offers the strongest limits. This likely results from Gemini-S having a larger collecting area than SPIRou's CFHT and the wavelength coverage (1.45-2.45 $\mu$m) of IGRINS extending further into the infrared than that of IRD (0.97-1.75 $\mu$m).

Overall, these upper limits indicate that GJ 486b lacks a clear H/He-dominated atmosphere with solar abundances.  However, some possible cloudy atmospheres or high-MMW secondary atmospheres are allowed by our data.

\section{Clouds and Hazes}

Clouds and hazes may be present in a wide range of exoplanet atmospheres \citep[e.g.,][]{Helling2019a} and tend to suppress spectral features in exoplanet transmission spectra \citep[e.g.,][]{Sing2016, Gandhi2020}.  As our models did not account for clouds (see Section \ref{sec:model_spectra}), the constraints presented in Section \ref{sec:InjectionAndRecoveryTests} are only valid under the assumption that GJ 486b's atmosphere is clear (i.e., free of clouds and haze).

\begin{figure*}
\centering
\begin{tabular}{cc}
  \includegraphics[width=0.5\textwidth,page=1]{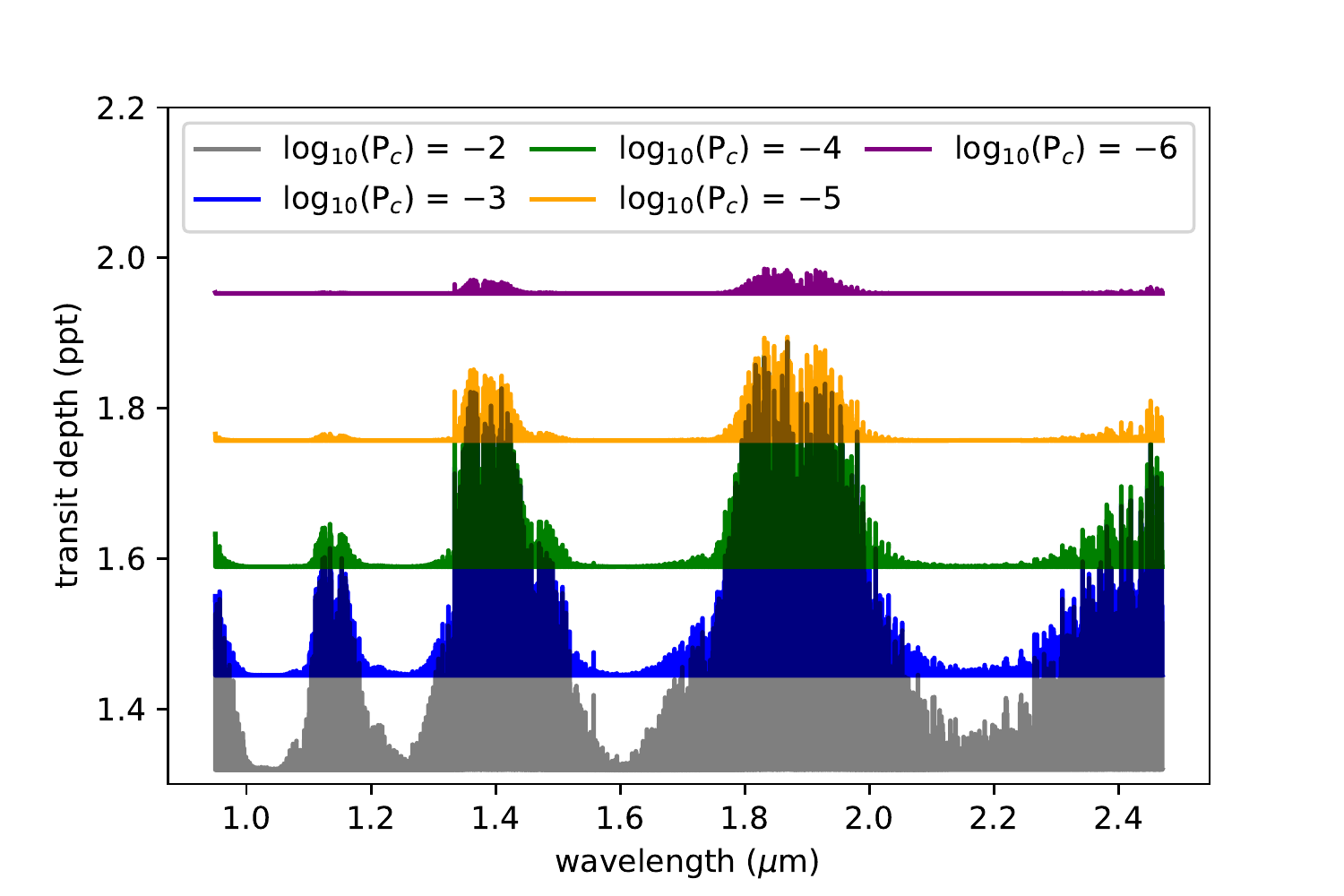} & \includegraphics[width=0.5\textwidth,page=2]{GJ486b_clouds.pdf} \\ \includegraphics[width=0.5\textwidth,page=3]{GJ486b_clouds.pdf} &
  \includegraphics[width=0.5\textwidth,page=4]{GJ486b_clouds.pdf} \\
  \end{tabular}
\caption{The effect of cloud decks on GJ 486b's spectrum. \textbf{Top left:} GJ 486b model spectra of a hydrogen-dominated atmosphere with water at solar abundances. The pressure of the cloud deck, $P_c$, is indicated in the legend as $\log_{10}(P_c$) and corresponds to a range of 10$^{-6}$ bar to 10$^{-2}$ bar. \textbf{Top right:} The same model spectra as in the top left, except the continuum has been subtracted.  \textbf{Lower left:} Auto-cross-correlation functions for the models shown in the top row. \textbf{Lower right:} Auto-cross-correlation peak amplitude.}
\label{fig:clouds}
\end{figure*}

To assess how our results would change without this assumption, we examined the effect of adding gray cloud decks at varying pressure levels to our hydrogen-dominated model with water at solar abundances (see Section \ref{sec:model_spectra}). These models have cloud decks at pressures of 10$^{-6}$ bar to 10$^{-2}$ bar with factors of 10 between steps.  As all the models used in this study set the planet radius to correspond to a pressure of 10$^{-2}$ bar, the model with a cloud deck at this level is essentially cloud-free.  These models are shown before and after continuum subtraction in the top left and right panels of Fig. \ref{fig:clouds}.  We find that lower cloud deck pressures (corresponding to higher altitudes) result in more suppression of the spectral lines. To quantify the level of suppression caused by the clouds, we calculated auto-cross-correlation functions for each model. The lower left and right panels of Fig. \ref{fig:clouds} show the auto-cross-correlation functions and peak auto-cross-correlation amplitude, respectively, after normalizing so that the maximum value is 1.  By comparing to the reference model with a cloud deck at 10$^{-2}$ bar (essentially cloud-free), we find that cloud decks at pressures of $\lesssim$10$^{-3}$ bar reduce the strength of the cross-correlation signal by a factor of $\gtrsim$3.  This implies that, if an injection/recovery test were performed (as in Section \ref{sec:InjectionAndRecoveryTests}) with a cloudy model, it would be recovered to a lower significance than the same model without clouds.  Therefore, some of the VMR-MMW parameter space that was ruled out by our data while assuming a clear atmosphere will still be allowed for a cloudy atmosphere.

\section{Implications for JWST \label{sec:JWST}}

GJ 486b will be observed by JWST in both transit and eclipse. Two transits of GJ 486b are scheduled to be observed with JWST's Near-InfraRed Spectrograph (NIRSpec), with the G395H grism in Bright Object Time Series (BOTS) mode, as part of a larger program to search for atmospheres on planets transiting M-dwarf stars (GO 1981\footnote{\url{https://www.stsci.edu/jwst/science-execution/program-information.html?id=1981}}, P.I.: K. Stevenson). Two eclipses with the Mid-Infrared Instrument (MIRI) will further probe GJ 486b's dayside atmosphere (GO 1743\footnote{\url{https://www.stsci.edu/jwst/science-execution/program-information.html?id=1743}}, P. I.: M. Mansfield).

Our constraints on GJ 486b's atmospheric composition (Section \ref{sec:InjectionAndRecoveryTests}) can inform these future JWST observations.  \citet{Caballero2022} show that a H$_2$/He-dominated atmosphere can be readily detected by JWST with two transits. However, we can confidently rule out to 5$\sigma$ a clear H$_2$/He-dominated atmosphere with H$_2$O, CH$_4$, NH$_3$, HCN, CO$_2$, and CO at solar abundances.

We further investigated the prospects for JWST to detect GJ 486b's atmosphere via simulated observations. We used PandExo \citep{Batalha2017} to simulate two transits of GJ 486b with NIRSpec G395H. To match the scheduled observations, we used an observation duration of 5.14 hr, the $R$ = 2700 f290lp grism, and the S1600A1 SUB2048 subarray option. We set the number of groups per integration to the `optimized' default, the saturation limit to 80\% of the full well capacity, and assumed no noise floor.

\begin{figure*}
\centering
\begin{tabular}{cc}
  \includegraphics[width=0.35\textwidth]{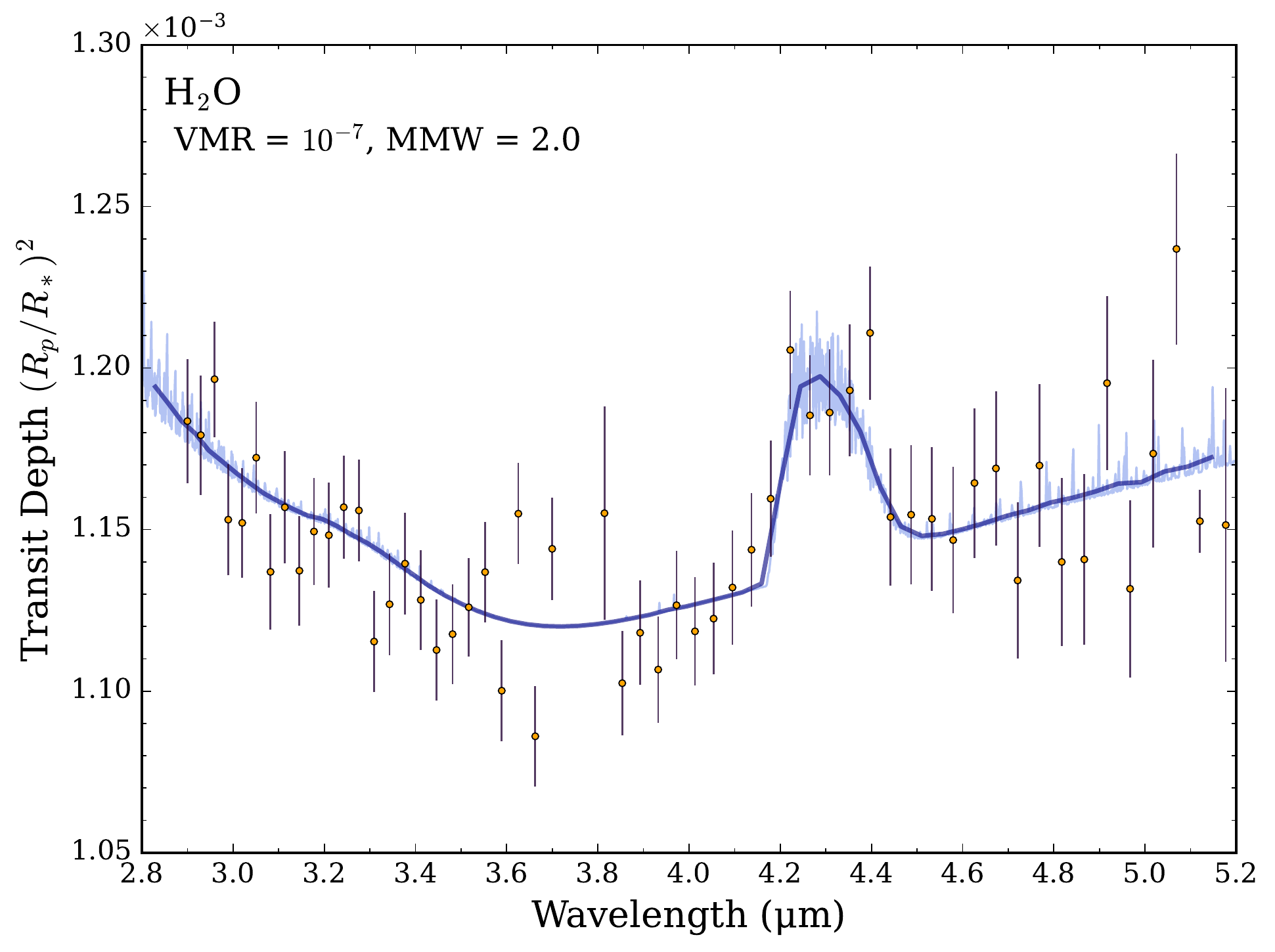} & \includegraphics[width=0.35\textwidth]{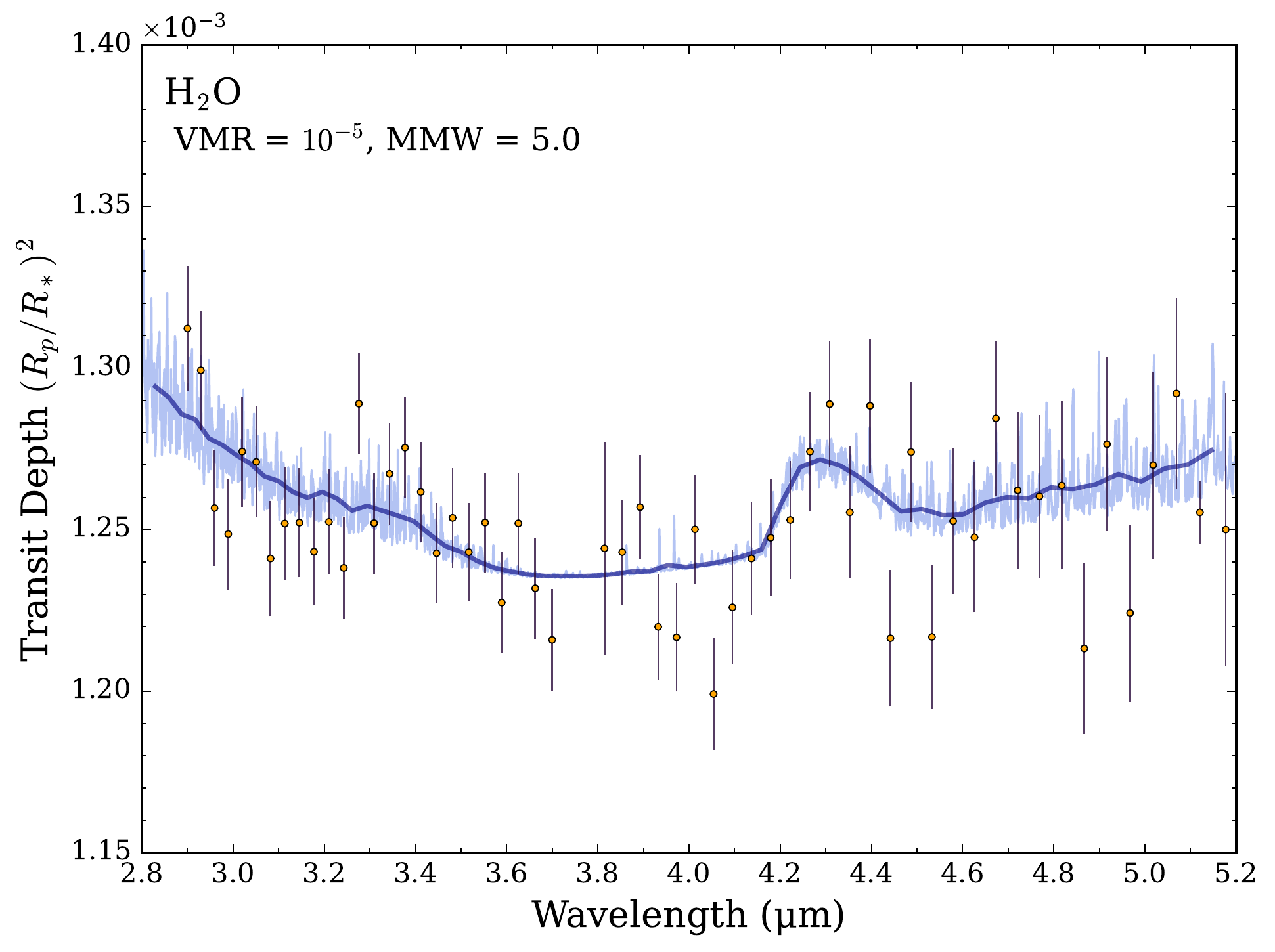} \\ \includegraphics[width=0.35\textwidth]{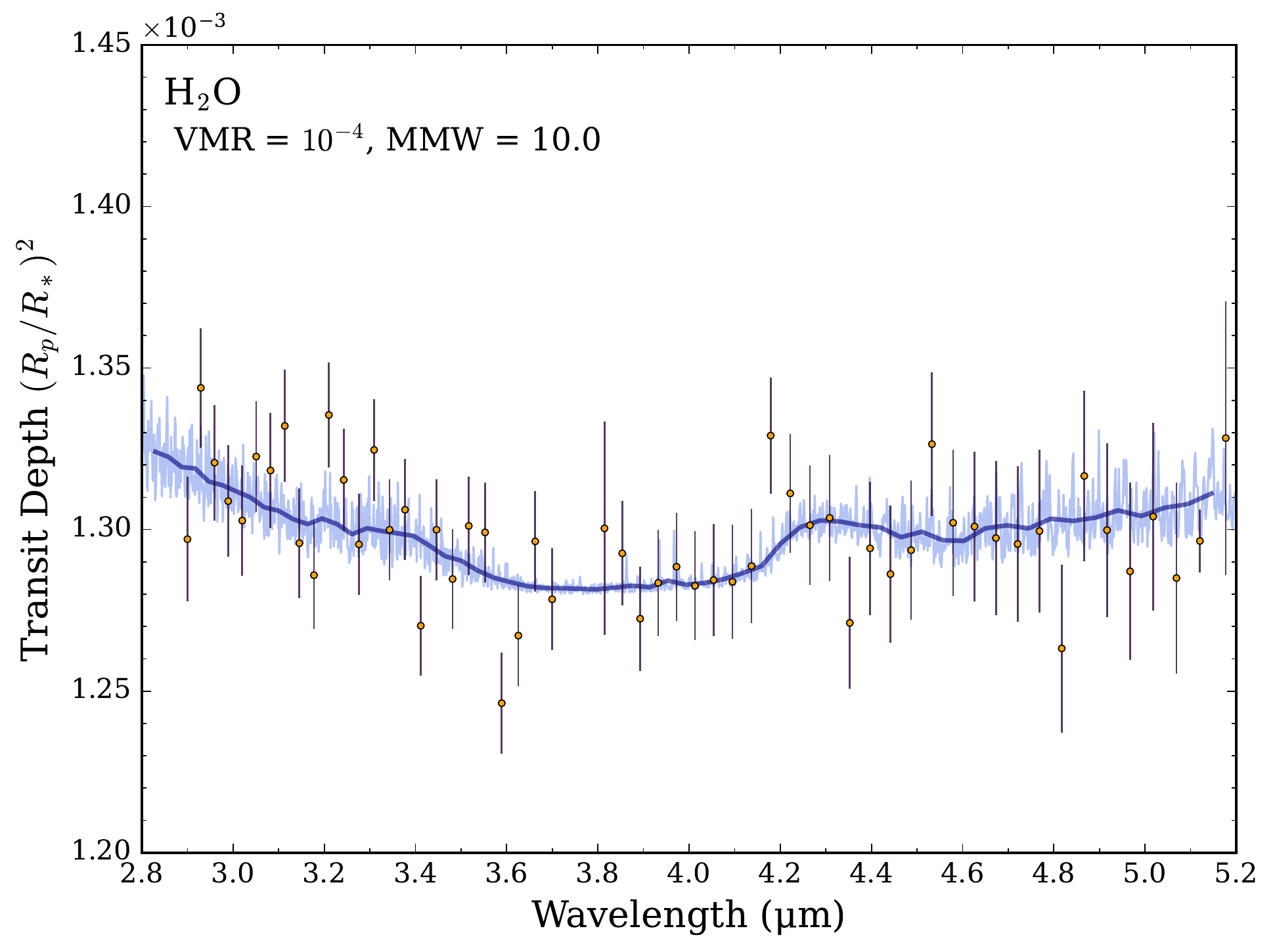} &
  \includegraphics[width=0.35\textwidth]{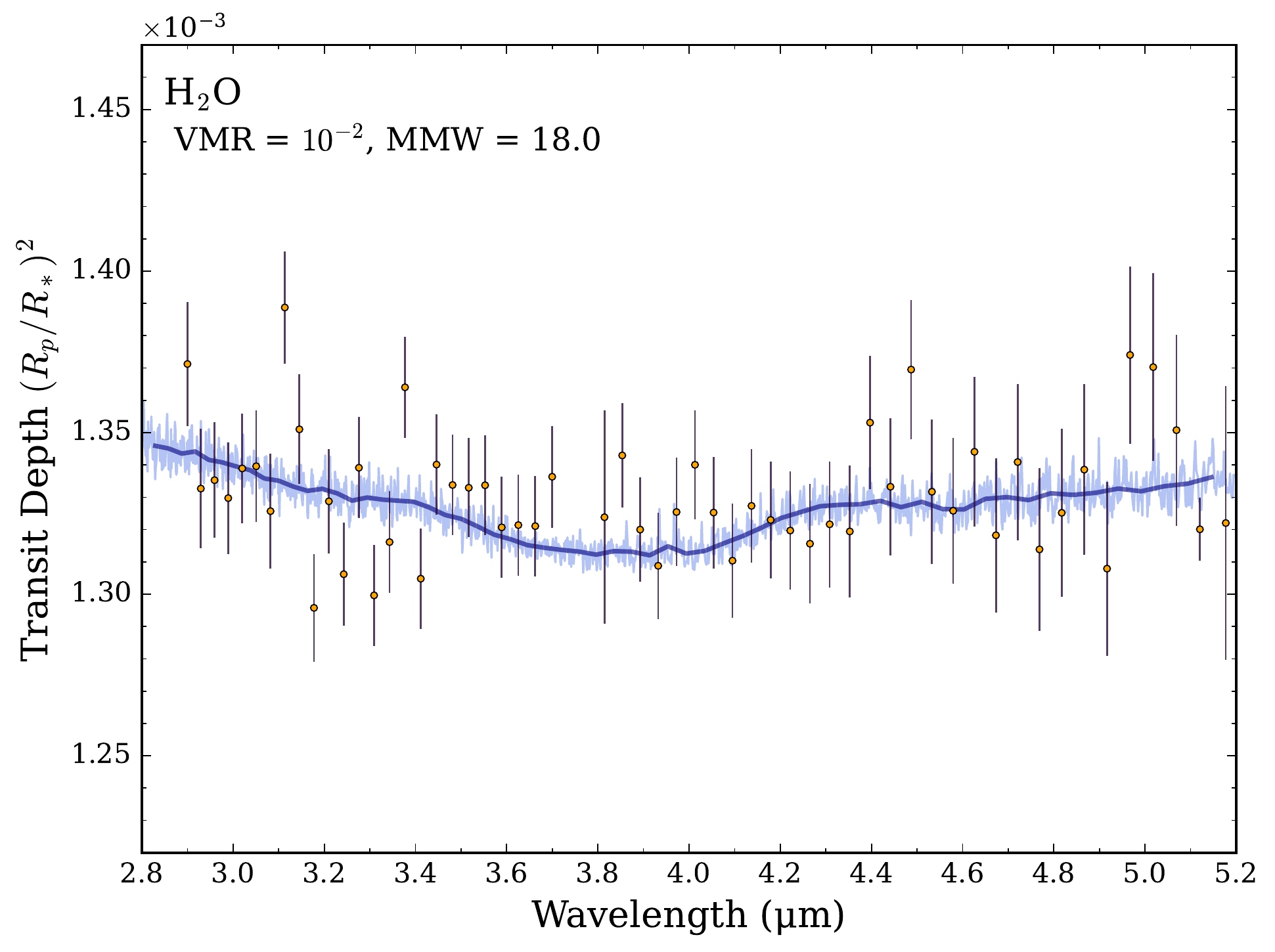} \\
    \includegraphics[width=0.35\textwidth]{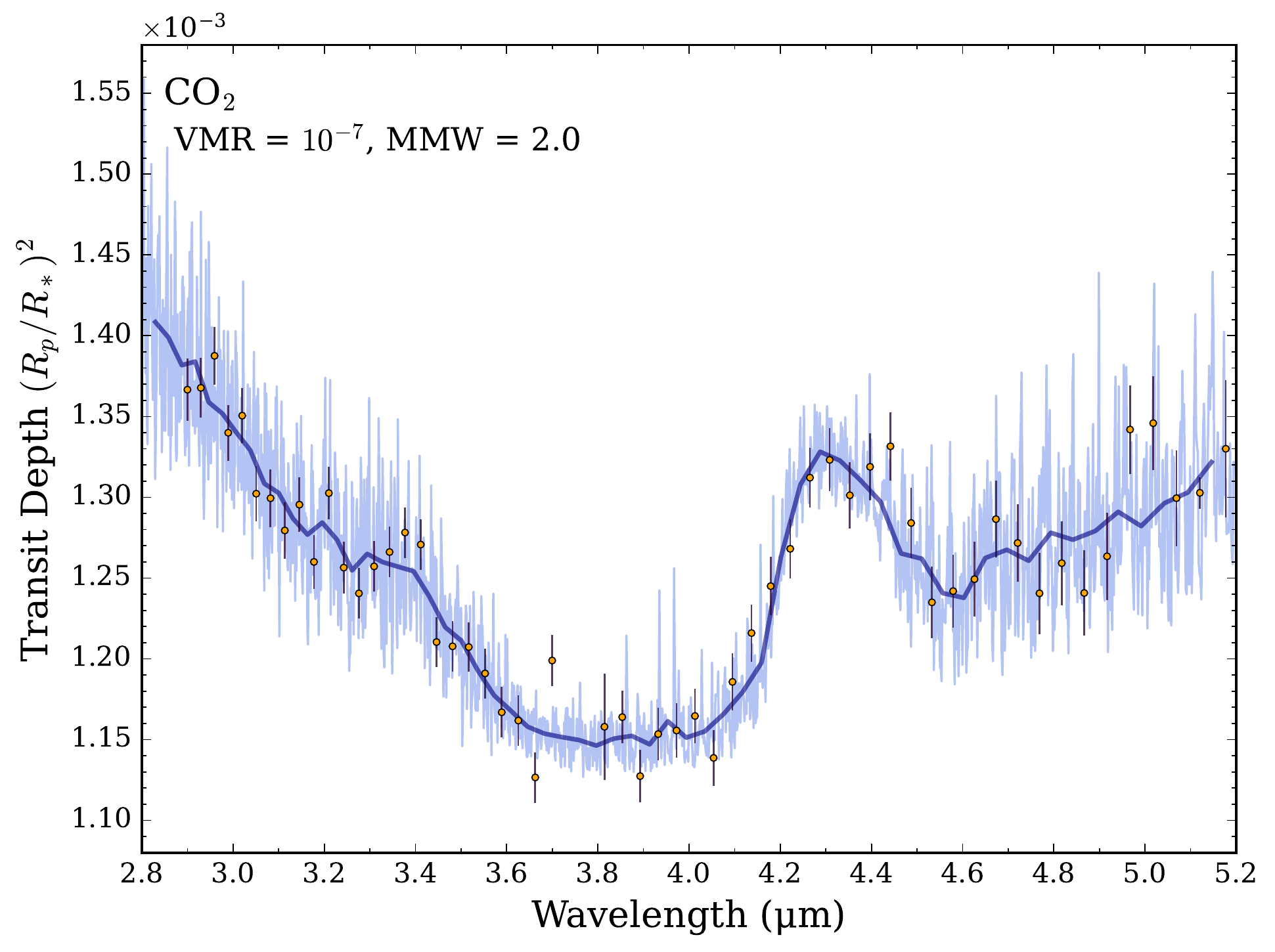} & \includegraphics[width=0.35\textwidth]{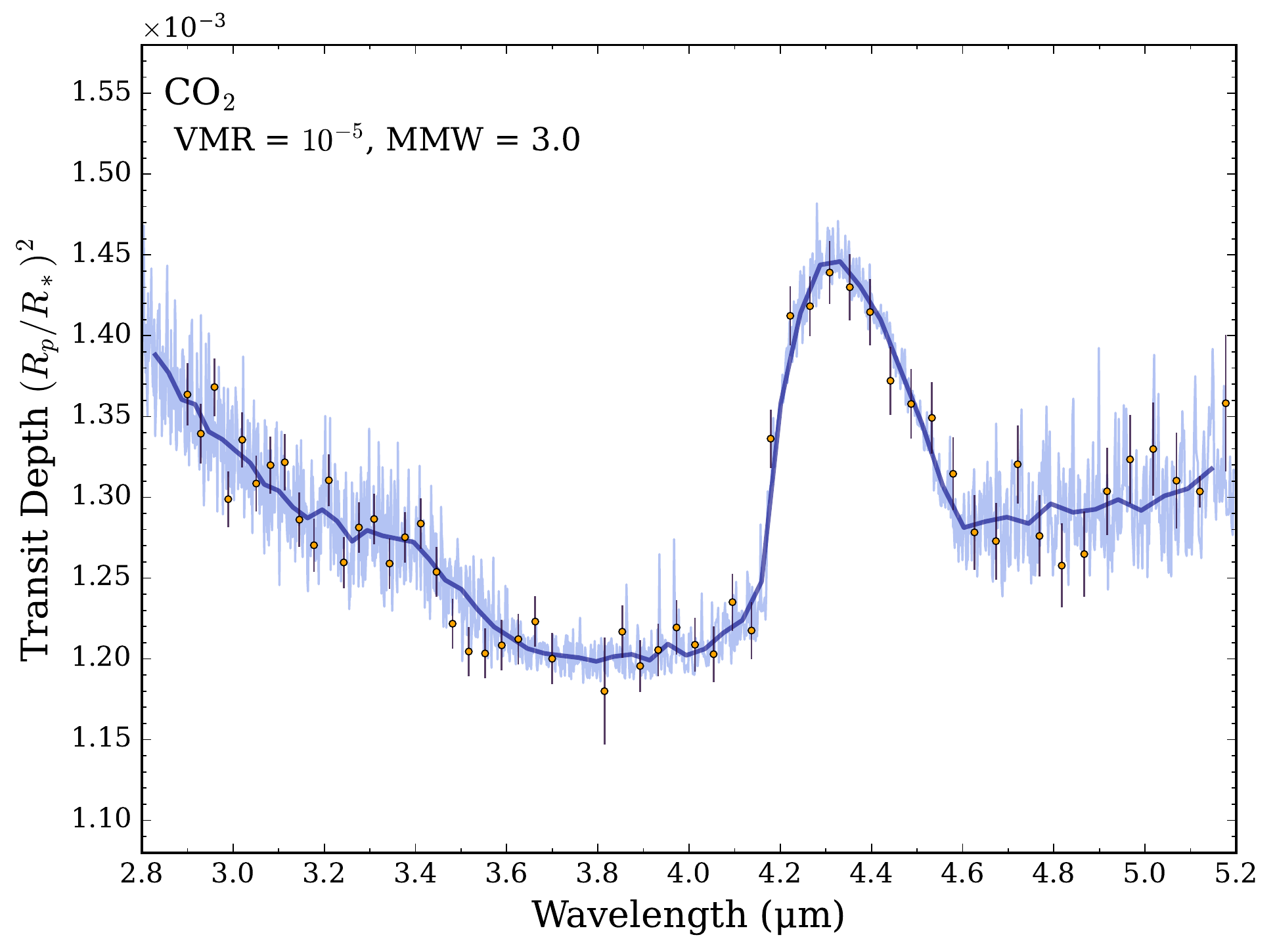} \\ \includegraphics[width=0.35\textwidth]{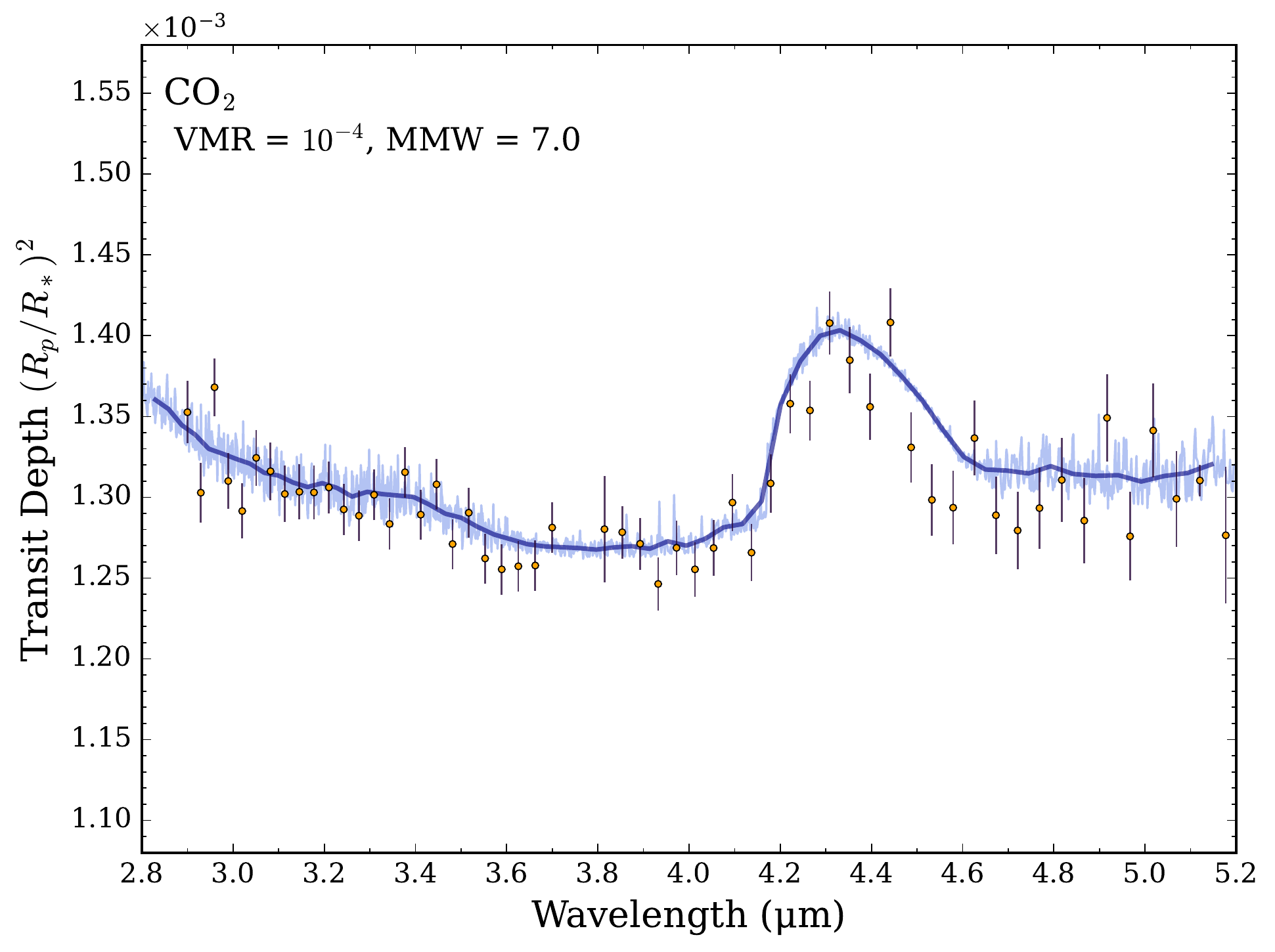} &
  \includegraphics[width=0.35\textwidth]{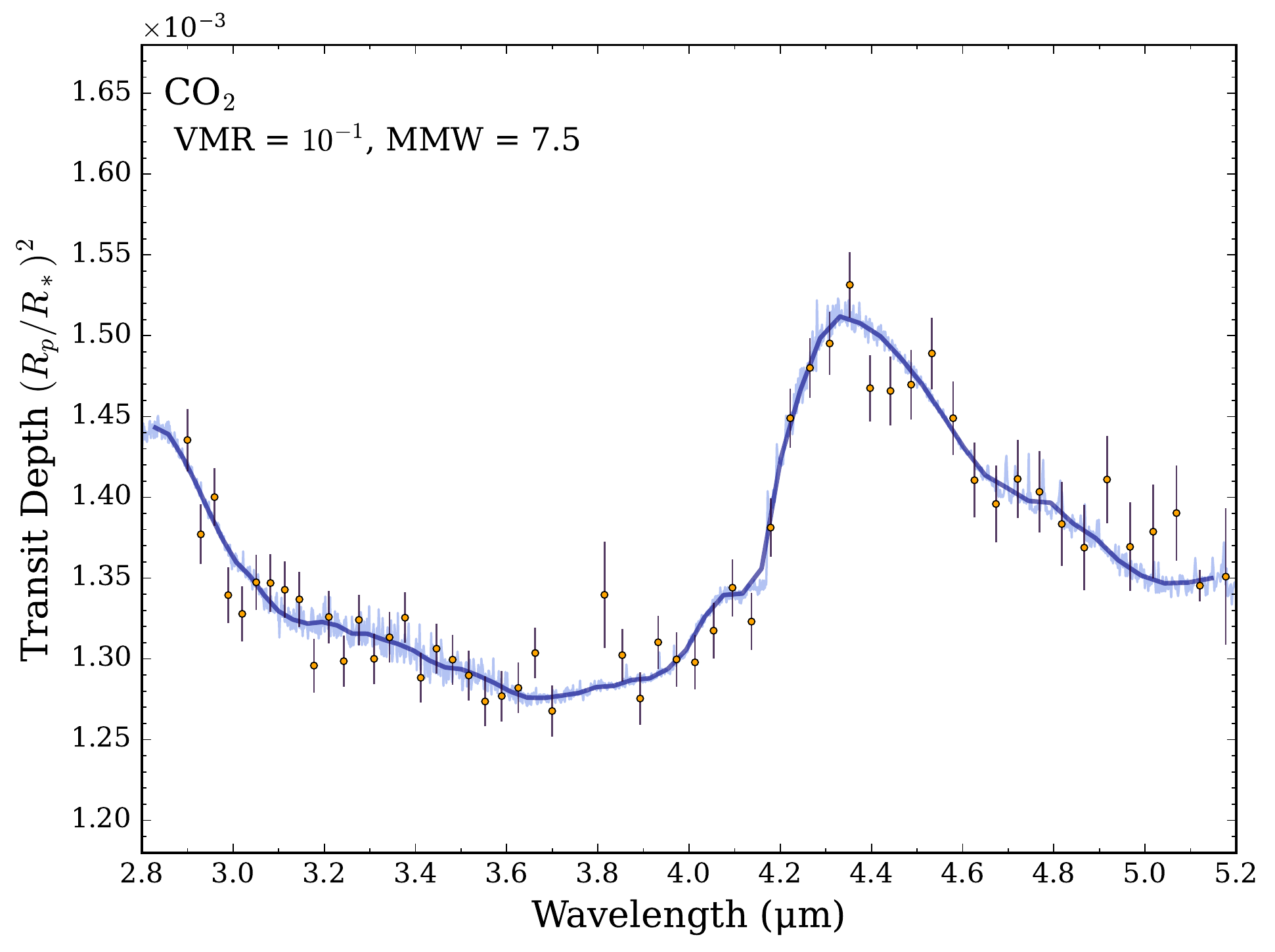} \\
  \end{tabular}
\caption{Simulated JWST transmission spectra for possible GJ 486b atmospheric compositions allowed by our high-resolution observations. In all panels, we show the models at a resolving power of $R$ = 2700 (light blue), corresponding to the native resolution of NIRSpec/G395H, as well as the models binned to a resolution of $R$ = 100 (navy lines). The top four panels show models with variable H$_2$O abundances and mean molecular weights with CO$_2$ fixed to a solar abundance. The lower four panels switch the roles of CO$_2$ and H$_2$O.}
\label{fig:PandExoModels}
\end{figure*}

We considered eight cloud-free model atmospheres compatible with our high-resolution observations.  We focused on H$_2$O and CO$_2$ as they are the main expected absorbers over the NIRSpec G395H spectral range. The models also contained background H$_2$. Four of the models vary the H$_2$O abundance and atmospheric mean molecular weight while CO$_2$ is fixed at a solar abundance (VMR$_{\sun, CO_2}$ $\approx$ 5$\times$10$^{-9}$). The other four models vary the CO$_2$ abundance and the atmospheric mean molecular weight while H$_2$O is fixed to a solar abundance (VMR$_{\sun, H{_2}O}$ $\approx$ 5$\times$10$^{-4}$). The models with variable H$_2$O have $\log_{10}$(VMR$_{H2O}$) = $-$7, $-$5, $-$4 and $-$2, and MMWs of 2, 5, 10, and 18, respectively. The models with variable CO$_2$ have $\log_{10}$(VMR$_{CO2}$) = $-$7, $-$5, $-$4 and $-$1, and MMWs of 2, 3, 7, and 7.5, respectively.  These values of VMR and MMW were chosen because they lie just outside the region of VMR-MMW parameter space ruled out by our high-resolution observations in Fig \ref{fig:CombinedConstraints}. The simulated JWST observations for these models are shown in Fig. \ref{fig:PandExoModels}. By comparing the simulated observation error bars to the size of the features in the model spectra, it is apparent that the scheduled transit observations could readily distinguish all considered models with variable CO$_2$ abundance and a solar H$_2$O abundance (lower four panels in Fig. \ref{fig:PandExoModels}). However, these observations may not be able to distinguish our considered models with variable H$_2$O abundance and a solar CO$_2$ abundance (upper four panels in Fig. \ref{fig:PandExoModels}).  This indicates the complementary capabilities of high-resolution ground-based and JWST observations, because the former has greater sensitivity to H$_2$O while the latter has greater sensitivity to CO$_2$. 

We note that our atmospheric models are unable to make predictions that could inform the eclipse observations --- since they have isothermal pressure-temperature profiles, which produce featureless emission spectra. Future investigations that also consider a range of possible pressure-temperature profiles may provide useful insights into GJ 486b's emission spectrum.  

Overall, these results indicate that JWST is likely to constrain terrestrial exoplanet atmospheres more strongly than the high-resolution ground-based instrumentation used in this study.  However, such ground-based instrumentation will still make important contributions to our understanding of terrestrial exoplanet atmospheres.  For example, the combination of high- and low-resolution spectroscopy can provide improved constraints over what would be possible with either method on its own \citep[e.g.,][]{Brogi2017, Khalafinejad2021}.  Furthermore, high-resolution spectroscopy is sensitive to line cores that probe altitudes above cloud decks \citep[e.g.][]{Gandhi2020}, so it can provide valuable insight into the atmospheres of cloudy terrestrial exoplanets.  High-resolution spectroscopy will play an even larger role in the characterization of terrestrial exoplanet atmospheres when the upcoming extremely large telescopes are operational. For example, it has been estimated that a high-dispersion spectrograph on the Extremely Large Telescope (ELT) may be an order of magnitude more sensitive than JWST to oxygen in a terrestrial exoplanet's atmosphere \citep{Snellen2013}. In this regime of high signal-to-noise ratios for terrestrial exoplanets, high-resolution spectroscopy's unique capabilities will allow detections of Doppler shifts \citep[which can indicate atmospheric winds;][]{Snellen2010}, line broadening \citep[which can indicate rotation;][]{Snellen2014} and potentially molecular isotopologues \citep{Molliere2019a}.

\section{Conclusion} \label{sec:conclusion}

In this study, we searched for the atmosphere of the 1.3 $R_\earth$ exoplanet GJ 486b using high-resolution transmission spectroscopy. Our resultant atmospheric constraints additionally offer an improved understanding of the possible atmospheres for terrestrial planets orbiting M-dwarf stars. 

We observed three transits of GJ 486b with the spectrographs IRD, IGRINS, and SPIRou, and searched for absorption by its atmosphere with the high-resolution cross-correlation method.  We searched for C$_2$H$_2$, CH$_4$, CO, CO$_2$, FeH, H$_2$O, HCN, H$_2$S, K, Na, NH$_3$, PH$_3$, SiO, TiO, and VO, but did not detect any robust atmospheric signals. Our best potential detection was H$_2$O at the 3$\sigma$ level, seen only in the IGRINS data set. However, this potential signal is likely an artifact of imperfectly removed H$_2$O lines in the stellar spectrum. We suggest that it may be possible to improve the suppression of these stellar lines by using 3D models of stellar convection, as shown by \citet{Chiavassa2019}.   

Nevertheless, we derived informative upper limits on the abundances of many chemical species by performing signal injection and recovery tests. Our recovery tests allow us to rule out a clear H$_2$/He-dominated atmosphere with solar molecular abundances to a confidence of $>$5$\sigma$, while also ruling out a clear 100\% water atmosphere to a confidence of $3\sigma$. Since a water-dominated atmosphere is often considered to be the most likely possibility for 3 $M_\earth$ planets like GJ 486b \citep[e.g.,][]{Ortenzi2020}, our results could have important implications for planetary interior models. 

We also investigated the implications of our findings for the upcoming JWST transit observations of GJ 486b. We found that these observations will be especially sensitive to the presence of CO$_2$, but with weaker sensitivity to H$_2$O, highlighting the complementary capabilities of high-resolution and JWST observations.

In summary, our results indicate that GJ 486b does not possess a cloud-free H/He-dominated atmosphere that has solar abundances. However, it may possess a secondary atmosphere with high mean molecular weight or a H$_2$/He-dominated atmosphere with clouds. Our findings provide further evidence suggesting that terrestrial planets orbiting M-dwarf stars experience significant atmospheric loss. 



\section*{Acknowledgments}
We thank the anonymous reviewer for constructive comments. We are also grateful to Hajime Kawahara for helpful suggestions.

This research is based [in part] on data collected at the Subaru Telescope, which is operated by the National Astronomical Observatory of Japan. We are honored and grateful for the opportunity of observing the Universe from Maunakea, which has cultural, historical, and natural significance in Hawaii.

This work used the Immersion Grating Infrared Spectrometer (IGRINS) that was developed under a collaboration between the University of Texas at Austin and the Korea Astronomy and Space Science Institute (KASI) with the financial support of the Mt. Cuba Astronomical Foundation, of the US National Science Foundation under grants AST-1229522 and AST-1702267, of the McDonald Observatory of the University of Texas at Austin, of the Korean GMT Project of KASI, and of the Gemini Observatory.

Based on observations obtained at the international Gemini Observatory, a program of NSF’s NOIRLab, which is managed by the Association of Universities for Research in Astronomy (AURA) under a cooperative agreement with the National Science Foundation on behalf of the Gemini Observatory partnership: the National Science Foundation (United States), the National Research Council (Canada), Agencia Nacional de Investigaci\'{o}n y Desarrollo (Chile), Ministerio de Ciencia, Tecnolog\'{i}a e Innovaci\'{o}n (Argentina), Minist\'{e}rio da Ci\^{e}ncia, Tecnologia, Inova\c{c}\~{o}es e Comunica\c{c}\~{o}es (Brazil), and Korea Astronomy and Space Science Institute (Republic of Korea).

Based on observations obtained at the Canada-France-Hawaii Telescope (CFHT), which is operated from the summit of Maunakea by the National Research Council of Canada, the Institut National des Sciences de l'Univers of the Centre National de la Recherche Scientifique of France, and the University of Hawaii. The observations at the Canada-France-Hawaii Telescope were performed with care and respect from the summit of Maunakea which is a significant cultural and historic site.  

Based on observations obtained with SPIRou, an international project led by Institut de Recherche en Astrophysique et Planétologie, Toulouse, France.

This research has made use of the Extrasolar Planet Encyclopaedia, NASA's Astrophysics Data System Bibliographic Services, and the NASA Exoplanet Archive, which is operated by the California Institute of Technology, under contract with the National Aeronautics and Space Administration under the Exoplanet Exploration Program.

M.T. is supported by JSPS KAKENHI grant Nos. 18H05442, 15H02063, and 22000005. S. K. N. is supported by JSPS KAKENHI grant No. 22K14092.

J.D.T's support for this work was provided by NASA through the NASA Hubble Fellowship grant $\#$HST-HF2-51495.001-A awarded by the Space Telescope Science Institute, which is operated by the Association of Universities for Research in Astronomy, Incorporated, under NASA contract NAS5- 26555.



%

\vspace{5mm}
\facilities{8.2 m Subaru telescope (IRD), 8.1 m Gemini-South telescope (IGRINS), 3.6 m Canada France Hawai’i Telescope (SPIRou)}


\software{\texttt{Astropy} \citep{Astropy2013, Astropy2018}; \texttt{SciPy} \citep{2020SciPy-NMeth}; \texttt{NumPy} \citep{harris2020array}; \texttt{petitRADTRANS} \citep{Molliere2019b}; \texttt{SpectRes} \citep{Carnall2017}.
          }


\newpage
\appendix


\section{Excluded wavelength regions \label{sec:Excluded_wavelength_regions}}

Here, we explicitly define the wavelength regions that were excluded from our analysis for having severe telluric contamination or insufficient signal-to-noise ratios (especially at the edges of the spectral orders). The excluded regions for the IGRINS data are shown in Table \ref{tab:ExcludedWavesForIGRINS}.


\begin{center}
\begin{longtable}{c c c c}
\caption{The IGRINS Orders Used to Measure the Drift in Its Initial Wavelength Calibration.} \label{tab:ExcludedWavesForIGRINS} \\

\hline
IGRINS order & Order used in  & \multicolumn{2}{c}{Number of edge pixels excluded} \\
wavelength ($\mu$m) & drift correction & left edge & right edge \\ \hline
\endfirsthead

\multicolumn{4}{c}%
{{\bfseries \tablename\ \thetable{} -- continued from previous page}} \\ 
\hline
IGRINS order & Order used in   & \multicolumn{2}{c}{Number of edge pixels excluded} \\
wavelength ($\mu$m) & drift correction & left edge & right edge \\ \hline  
\endhead

\hline \multicolumn{4}{|r|}{{Continued on next page}} \\ \hline
\endfoot

\hline \hline
\endlastfoot

\hline
1.429 $-$ 1.449 & no & \multicolumn{2}{c}{entire order excluded} \\
1.440 $-$ 1.460 & no & \multicolumn{2}{c}{entire order excluded} \\
1.452 $-$ 1.472 & no & 552 & 107 \\
1.463 $-$ 1.483 & no & 552 & 105 \\
1.475 $-$ 1.495 & yes & 526 & 105 \\
1.486 $-$ 1.507 & yes & 494 & 105 \\
1.498 $-$ 1.519 & yes & 461 & 105 \\
1.511 $-$ 1.532 & yes & 445 & 105 \\
1.523 $-$ 1.544 & no & 433 & 114 \\
1.536 $-$ 1.557 & no & 426 & 105 \\
1.549 $-$ 1.570 & no & 418 & 105 \\
1.562 $-$ 1.584 & yes & 406 & 105 \\
1.575 $-$ 1.597 & no & 411 & 105 \\
1.589 $-$ 1.611 & yes & 397 & 105 \\
1.603 $-$ 1.625 & no & 392 & 105 \\
1.617 $-$ 1.639 & no & 390 & 105 \\
1.631 $-$ 1.654 & yes & 390 & 105 \\
1.646 $-$ 1.669 & no & 390 & 105 \\
1.661 $-$ 1.684 & no & 390 & 105 \\
1.676 $-$ 1.699 & yes & 388 & 105 \\
1.691 $-$ 1.715 & no & 239 & 105 \\
1.707 $-$ 1.731 & no & 238 & 105 \\
1.723 $-$ 1.747 & no & 238 & 105 \\
1.740 $-$ 1.764 & no & 239 & 105 \\
1.756 $-$ 1.781 & no & 237 & 109 \\
1.774 $-$ 1.798 & no & 238 & 107 \\
1.791 $-$ 1.816 & no & 256 & 183 \\
1.809 $-$ 1.834 & no & \multicolumn{2}{c}{entire order excluded} \\
1.847 $-$ 1.873 & no & \multicolumn{2}{c}{entire order excluded} \\
1.866 $-$ 1.892 & no & \multicolumn{2}{c}{entire order excluded} \\
1.885 $-$ 1.911 & no & \multicolumn{2}{c}{entire order excluded} \\
1.905 $-$ 1.931 & no & \multicolumn{2}{c}{entire order excluded} \\
1.925 $-$ 1.952 & no & 336 & 162 \\
1.946 $-$ 1.973 & no & 265 & 148 \\
1.967 $-$ 1.994 & no & 240 & 140 \\
1.989 $-$ 2.016 & no & 229 & 172 \\
2.011 $-$ 2.039 & no & 260 & 118 \\
2.033 $-$ 2.062 & no & 220 & 112 \\
2.057 $-$ 2.085 & no & 213 & 105 \\
2.080 $-$ 2.109 & no & 207 & 105 \\
2.105 $-$ 2.134 & yes & 202 & 105 \\
2.130 $-$ 2.159 & no & 196 & 105 \\
2.155 $-$ 2.185 & yes & 190 & 105 \\
2.181 $-$ 2.212 & no & 183 & 105 \\
2.208 $-$ 2.239 & yes & 176 & 105 \\
2.236 $-$ 2.267 & yes & 170 & 105 \\
2.264 $-$ 2.295 & yes & 161 & 105 \\
2.293 $-$ 2.325 & no & 155 & 105 \\
2.323 $-$ 2.355 & yes & 155 & 105 \\
2.353 $-$ 2.386 & no & 155 & 105 \\
2.384 $-$ 2.417 & no & 155 & 105 \\
2.417 $-$ 2.450 & no & 160 & 111 \\
2.450 $-$ 2.484 & no & 167 & 242 \\
2.484 $-$ 2.518 & no & \multicolumn{2}{c}{entire order excluded} \\
\hline
\multicolumn{4}{l}{Notes. Also shown are the wavelength regions of each IGRINS} \\
\multicolumn{4}{l}{order that were excluded from the analysis due to severe telluric} \\
\multicolumn{4}{l}{contamination or low signal-to-noise ratio. All orders initially} \\
\multicolumn{4}{l}{have 2048 pixels.} \\
\end{longtable}

\end{center}







\bibliography{references}{}
\bibliographystyle{aasjournal}



\end{document}